\documentclass[12pt,preprint]{aastex}

\newcommand{\kmpers}{{{\rm \; km\;s}^{-1}}}

\received{}
\revised{}
\accepted{}

\shorttitle{Exoplanetary Atmospheres Survey}

\shortauthors{Jensen et al.}

\begin{document}

\title{A SURVEY OF ALKALI LINE ABSORPTION IN EXOPLANETARY ATMOSPHERES}

\author{Adam G.~Jensen\altaffilmark{1}, Seth Redfield\altaffilmark{1}, Michael Endl\altaffilmark{2}, William D.~Cochran\altaffilmark{2}, Lars Koesterke\altaffilmark{2,3}, \& Travis Barman\altaffilmark{4}}

\altaffiltext{1}{Van Vleck Observatory, Astronomy Department, Wesleyan University, 96 Foss Hill Drive, Middletown, CT 06459; Adam.Jensen@gmail.com, sredfield@wesleyan.edu}
\altaffiltext{2}{University of Texas, Department of Astronomy, Austin, TX 78712; mike@astro.as.utexas.edu, wdc@astro.as.utexas.edu, lars@tacc.utexas.edu}
\altaffiltext{3}{Texas Advanced Computing Center, Research Office Complex 1.101, J.~J.~Pickle Research Campus, Building 196, 10100 Burnet Road (R8700), Austin, Texas 78758-4497}
\altaffiltext{4}{Lowell Observatory, 1400 West Mars Hill Road, Flagstaff, AZ 86001; barman@lowell.edu}

\begin{abstract}
We obtained over 90 hours of spectroscopic observations of four exoplanetary systems with the Hobby-Eberly Telescope (HET).  Observations were taken in transit and out of transit, and we analyzed the differenced spectra---i.e., the transmission spectra---to inspect it for absorption at the wavelengths of the neutral sodium (\ion{Na}{1}) doublet at $\lambda\lambda5889, 5895$ and neutral potassium (\ion{K}{1}) at $\lambda7698$.  We used the transmission spectrum at \ion{Ca}{1} $\lambda6122$---which shows strong stellar absorption but is not an alkali metal resonance line that we expect to show significant absorption in these atmospheres---as a control line to examine our measurements for systematic errors.  We use an empirical Monte Carlo method to quantity these systematic errors.  In a reanalysis of the same dataset using a reduction and analysis pipeline that was derived independently, we confirm the previously seen \ion{Na}{1} absorption in HD 189733b at a level of $(-5.26\pm1.69)\times10^{-4}$ (the average value over a 12 \AA{} integration band to be consistent with previous authors).  Additionally, we tentatively confirm the \ion{Na}{1} absorption seen in HD 209458b (independently by multiple authors) at a level of $(-2.63\pm0.81)\times10^{-4}$, though the interpretation is less clear.  Furthermore, we find \ion{Na}{1} absorption of $(-3.16\pm2.06)\times10^{-4}$ at $<3\sigma$ in HD 149026b; features apparent in the transmission spectrum are consistent with real absorption and indicate this may be a good target for future observations to confirm.  No other results (\ion{Na}{1} in HD 147506b and \ion{Ca}{1} and \ion{K}{1} in all four targets) are significant to $\geq 3\sigma$, although we observe some features that we argue are primarily artifacts.
\end{abstract}

\section{INTRODUCTION AND BACKGROUND}
\label{s:intro}
Astronomers have now had evidence of planets orbiting other stars for roughly two decades, including:  the possibility of a planet around HD 114762 described by \citet{Latham1989}; evidence of planets around pulsars due to timing variations \citep{WolszczanFrail1992, Rasio1994}; and the famous detection of a ``hot Jupiter" around the solar-type star 51 Peg \citep{MayorQueloz1995}.  The \citet{Latham1989} and \citep{MayorQueloz1995} discoveries used the radial velocity (RV) method, observing the cyclic Doppler shifts in a star's spectrum to detect its orbit around a center of mass caused by one or more unseen companions.  As of this writing, the RV method is still the dominant discovery method for confirmed detections of exoplanets.

Over the last decade or so, astronomers have also observed the effects of exoplanets that pass in front of their host stars and create measurable reductions in the observed light of those stars \citep{Charbonneau2000, Henry2000, Mazeh2000}.  Transits are an extremely important phenomenon because they uniquely provide critical characteristics of planetary systems, such as the ratio of the planetary and stellar radii, and the ratio of the planetary orbit and the stellar radius.  RV measurements, in contrast, cannot provide any information regarding the former, and can only be used to determine the latter if assumptions from stellar models are used.  Planets that transit their central stars also present the only practical opportunity for astronomers to observe the properties of exoplanetary atmospheres in transmission.  This is important because transmission spectroscopy of sufficient resolution is arguably the least ambiguous way to identify specific atomic or molecular species, thus providing important information about exoplanetary atmospheric composition.

Many of the exoplanets that are known are ``hot Jupiters," planets with masses approximately that of Jupiter and with orbital semi-major axis $a \ll 1$ AU.  Such planets are observationally favored by searches using both radial velocity and transit methods.  Consequently, these are also the planets most conducive to studies of their atmospheres using transmission spectroscopy.

There are many different dimensions of the detection of an exoplanetary atmosphere.  The central issue is the atomic or molecular species being detected.  Detecting an atmosphere requires a differential measurement to distinguish the atmosphere from the signal of the disk, and thus all observations can be broadly classified as spectroscopy, ranging from narrow-band photometry to high-resolution spectroscopy.  The wavelength, or more broadly the waveband (e.g.~UV, optical, IR), dictates properties of the observation such as the telescope, instrument, and whether the observation is ground- or space-based.  Detections may be made in absorption, using transits, or emission, using secondary eclipses and direct spectra.  Finally, detections may be of a bound atmosphere at a low number of characteristic scale heights or an unbound exosphere at many scale heights.  All of these different dimensions are intertwined, but practically result in a very heterogeneous set of atmospheric observations that have been made at this early stage in the history of exoplanetary study.  We discuss these results below.


\subsection{Overview of Existing Atmospheric Detections}
\label{ss:overview}
The literature contains a relatively small but rapidly increasing number of detections of a specific atomic or molecular species in an exoplanetary atmosphere.  Due to observational constraints, these detections have been primarily limited to only a few targets.  \citet{Charbonneau2002} was the first to detect an exoplanetary atmosphere.  Using optical spectroscopy from the Space Telescope Imaging Spectrograph (STIS) onboard the {\it Hubble Space Telescope} ({\it HST}), \citet{Charbonneau2002} detected neutral sodium in HD 209458b.  Soon after, \citet{VidalMadjar2003, VidalMadjar2004} discovered the exosphere (the extended, unbound portion of an atmosphere) of HD 209458b by detecting hydrogen, carbon, and oxygen using the ultraviolet capabilities of STIS; \citet{LecavelierDesEtangs2010} made a similar detection of hydrogen in HD 189733b using ACS spectroscopy.  In contrast to these space-based detections, \citet{Redfield2008} made the first ground-based detection of an exoplanetary atmosphere using the Hobby-Eberly Telescope (HET), detecting sodium in HD 189733b.  In the same year, \citet{Snellen2008} also made an important ground-based observation, using Subaru to confirm the \citet{Charbonneau2002} detection of sodium in HD 209458b.

The IR has also been an extremely fruitful waveband for detecting exoplanetary atmsopheres.  For example, there have been detections of water vapor (H$_2$O), carbon monoxide (CO), carbon dioxide (CO$_2$), and methane (CH$_4$) in HD 209458b and HD 189733b \citep{Barman2007, Beaulieu2008, Beaulieu2010, Grillmair2008, Swain2008, Swain2009a, Swain2009b, Swain2010, Tinetti2007}.  We note, however, that some of these detections that used Spitzer are in significant dispute.  For example, \citet{Ehrenreich2007} present an alternate interpretation of the Spitzer dataset used by \citet{Tinetti2007} to claim a detection of H$_2$O in HD 209458b.  \citet{Gibson2011} also dispute molecular detections in HD 189733b and XO-1b that were made using {\it HST}/NICMOS.

The above discussion clearly shows how the study of exoplanetary atmospheres has been focused on the two targets HD 189733b and HD 209458b.  These systems are nearby and therefore the apparent brightness of their central stars enables transmission spectroscopy at greater signal-to-noise levels.  More recently, however, there have been several additional detections of exoplanetary atmospheres.  These detections include: observations of sodium in WASP-17b \citep{Wood2011}; the first detections of potassium, in HD 80606b \citep{Colon2010} and XO-2b \citep{Sing2011}; and the presence of metals in the extended atmosphere of WASP-12b \citep{Fossati2010}.

\subsection{Alkali Metal Absorption in Exoplanetary Atmospheres}
\label{ss:alkaliresults}
With the detection of transiting exoplanets, early models \citep{SeagerSasselov2000, Barman2001, Brown2001, Hubbard2001} sought to predict what features might dominate the atmospheres of hot Jupiter-type exoplanets.  In the optical, the two dominant features should be the resonance absorption lines of the alkali metals sodium and potassium.  \citet{Charbonneau2002} sought to test this prediction by observing the Na D doublet at $\lambda\lambda5889,5895$.  They used STIS at a spectroscopic resolution of $R\equiv\lambda/\delta\lambda=5540$, and created a transmission spectrum of the atmosphere of HD 209458b by subtracting the coadded in-transit spectra from the coadded out-of-transit spectra.  \citet{Charbonneau2002} found an absorption level of $(2.32\pm0.57)\times10^{-4}$ integrated over a 12 \AA{} band relative to the adjacent spectral regions, with decreasing levels of absorption detected as the bandwidth was increased.  From the ground, \citet{Snellen2008} confirmed this result, finding a $(5.6\pm0.7)\times10^{-4}$ absorption level summed over two 3 \AA{} bands centered on the two lines of the Na D doublet, which \citet{Snellen2008} determined was consistent with the \citet{Charbonneau2002} result when differences in bandpasses are considered.  \citet{Sing2008a} reanalyzed the \citet{Charbonneau2002} data and calculated the absorption level over several different bandpasses.  They concluded that both narrow and broad absorption was present, at a level consistent with early cloudless models \citep{SeagerSasselov2000, Brown2001, Hubbard2001}.

\citet{Redfield2008} detected sodium in the atmosphere of HD 189733b, at a level of $(6.72\pm2.07)\times10^{-4}$.  This measurement is an integration of the transmission spectrum $S_T=(F_{in}/F_{out})-1$, where $F_{in}$ and $F_{out}$ are, respectively, the normalized in- and out-of-transits fluxes.  The absorption level detected by \citet{Redfield2008} was recently confirmed by {\it HST} measurements (D.~Sing, 2010, private communication).  \citet{Redfield2008} integrated $S_T$ over a 12 \AA{} bandpass for consistency with the \citet{Charbonneau2002} observations of HD 209458b.
We do not necessarily expect these atmospheres to be similar, as infrared observations indicate that the atmosphere of HD 209458b has a temperature inversion \citep{Knutson2008} but the atmosphere of HD 189733b does not \citep{Grillmair2007}.  It is important to note that \citet{Pont2008} found no distinct features using the Advanced Camera for Surveys (ACS) onboard {\it HST}, implying a broad haze that does not allow for an absorption signature.  However, the low spectral resolution \citep[$R\sim$100, compared to $R\sim$60,000 in][]{Redfield2008} might mask the absorption.

A third detection of sodium was made recently by \citet{Wood2011}, who found absorption in WASP-17b at a level of $(5.5\pm1.3)\times10^{-4}$ using a $2\times1.5$ \AA{} bin.  This is significantly less than predicted by taking the model of \citet{Brown2001} applied to HD 209458b and scaling it by the difference in scale height of the two planets.  Also notable is that the absorption is significant only using narrow bandpasses; as the bandpass is increased, the absorption disappears.  \citet{Wood2011} discuss several possibilities to explain the discrepancy between models and their observations, including photoionization on the day side of the planet, the depletion of heavy elements on the cold night side, and clouds.  In particular, clouds could explain the nature of the absorption as a function of bandpass width, as broad absorption should be formed at high pressures and low altitudes and narrower absorption at lower pressures and higher altitudes.  Thus, a layer of clouds or other haze at the proper height would suppress the broad absorption but not the narrower absorption.

Recently, the first detections of potassium were made for HD 80606b \citep{Colon2010} and XO-2b \citep{Sing2011}.  Both detections where made using tunable-filter photometry with the OSIRIS instrument on the Gran Telescopio Canarias.  \citet{Sing2011} find clear absorption in the \ion{K}{1} core at 7664.9 \AA{} and in the blue \ion{K}{1} wing at 7582 \AA{}.  They find the apparent planetary radius to be consistent with cloudless models for XO-2b.  In contrast, \citet{Colon2010} see broad absorption but no additional absorption in the \ion{K}{1} core, implying that low-pressure, high-altitude potassium has been ionized or has condensed.

Taken as a whole---and given that broader searches in the optical have not revealed any other significant features---these results confirm that neutral sodium and potassium provide the dominant line opacity sources in the optical, in accordance with models.  They also highlight the issues that still exist.  In a given planet, what is the effect of photoionization and condensation on abundances?  How common are clouds and haze?  New observations of additional targets, preferably at high resolution, will be required to fully address these questions.

\subsection{Future Prospects and This Study}
\label{ss:thisstudy}
The trends in the detection of exoplanetary atmospheres discussed in \S\ref{ss:overview} and \S\ref{ss:alkaliresults} show that we are entering an exciting time for this subfield.  A variety of techniques have now been demonstrated to be feasible (ground- vs.~spaced-based, photometric vs.~spectroscopic, etc.).  Having an array of techniques and an ever-increasing target list means we are on the cusp of performing significant comparative atmospheric exoplanetology for the first time.  It would be particularly useful to have a homogeneous survey of many different targets in order to compare absorption in exoplanetary atmospheres, and determine if variations are correlated with stellar or planetary characteristics.

The study described in this paper aims to be a first step toward such a survey.  We have made uniform observations of the spectra of four exoplanetary systems.  The goal is to examine, with high-resolution spectroscopy, the transmission spectra of several exoplanetary atmospheres and search for absorption from the \ion{Na}{1} doublet at $\lambda\lambda5889,5995$, along with the \ion{K}{1} absorption line at $\lambda7698$.\footnote{Note that the \ion{K}{1} line is also part of a widely-spaced doublet, but the other component at $\lambda7664$ is close to much stronger telluric absorption than the 7698 \AA{} line.}  We describe the observations and basic data reduction in \S\ref{s:obsdata}.  In \S\ref{s:analysis} we describe our procedures for analyzing the reduced spectra, along with a discussion of vetting our results, including error analysis and examining ``control lines" for comparison.  In \S\ref{s:results} we describe our basic results.  We discuss these results in \S\ref{s:discussion} and summarize in \S\ref{s:summary}.


\section{OBSERVATIONS AND DATA REDUCTION}
\label{s:obsdata}

\subsection{Description of Observations}
\label{ss:obsdescription}
Observations were obtained with the HET between August 2006 and November 2008.  The driving factor behind our use of the HET is its large primary mirror with an effective aperture of 9.2m.  This is large enough to provide the high signal-to-noise that we require in these observations.  Spectra were observed with the High-Resolution Spectrograph (HRS) at a nominal resolution of $R\sim60,000$, though we measure wavelength-dependent and time-variable resolutions at our specific wavelengths of interest (see \S\ref{ss:reduction}).  Individual exposures were limited to 10 minutes in length in order to avoid saturation, though some are shorter.  Approximately 20\% of the exposures for each target were taken while its respective planet was in transit.  Specific details are provided in Tables \ref{table:observations} (observational statistics) and \ref{table:systemparams} (system and orbital parameters).

We have an interest in spectral features that are near telluric lines and we also have very high signal-to-noise requirements.  Therefore, part of our observing strategy was to observe ``custom" telluric standard stars to use in removal of Earth's atmospheric lines.  These stars were chosen to be near our exoplanetary targets on the sky and the observations were taken either immediately before or after our primary science observations.  The spatial and temporal proximity reduces differences in air mass and water vapor content between the primary observations and the telluric observations.

\subsection{Data Reduction}
\label{ss:reduction}
\begin{figure}[t!]
\begin{center}
\epsscale{1.0}
\plotone{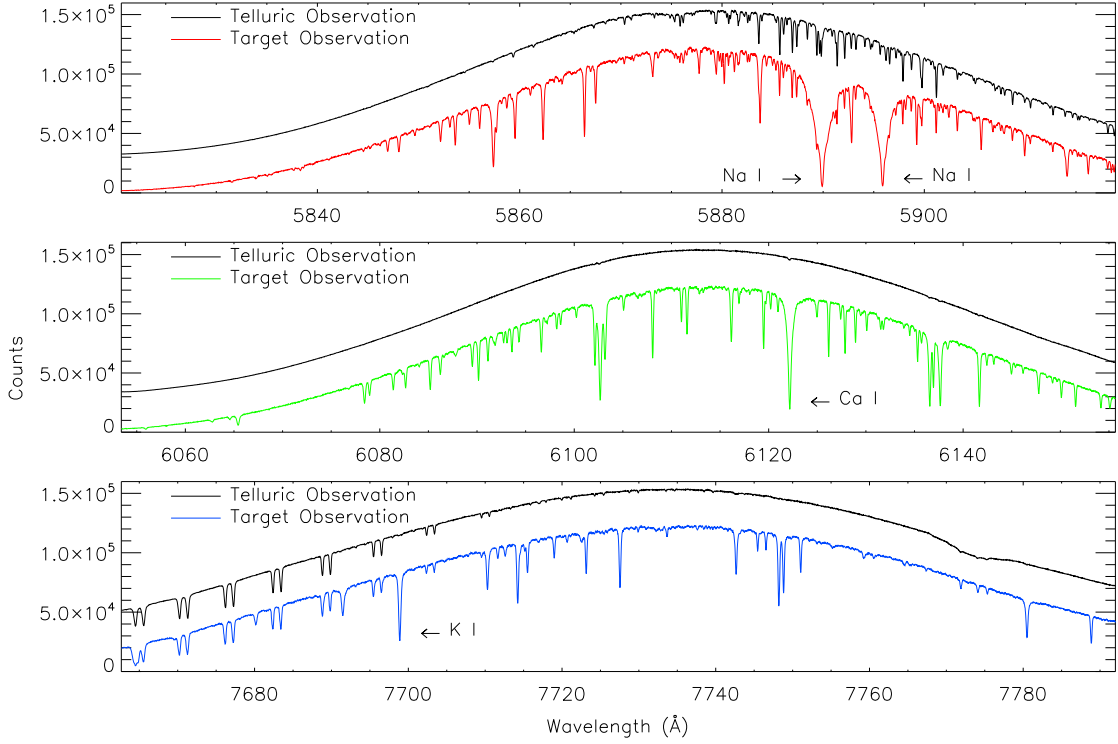}
\end{center}
\caption[]{Example of the spectra in the echelle orders containing \ion{Na}{1} $\lambda5889,5895$   \ion{Ca}{1} $\lambda6122$ \ion{K}{1} $\lambda7698$ for HD 189733, along with the spectra of the telluric standard star HD 196867.  All spectra are reduced from the raw echelle images but are not processed with the additional steps we discuss in \S\ref{ss:reduction} (including removal of stellar and ISM features from the telluric spectra, telluric subtraction, and normalization).  The target spectra are shown on the count scale of the y-axis, but the telluric spectra ars scaled to have the same maximum as the corresponding target spectra, and then arbitrarily shifted by 25\% of that same maximum.  The global shapes of the spectra are dominated by the blaze function of the echelle.  In the central panel, note the lack of telluric features in the telluric standard spectrum near \ion{Ca}{1}.  Also note that the line of interest is closer to the peak of the blaze function than either \ion{Na}{1} or \ion{K}{1}.  In the bottom panel, note the broad shallow feature near 7773 \AA{}.  While this feature does not affect our measurements at \ion{K}{1}, similar (but less dramatric) stellar and interstellar features are modeled and removed from the telluric spectrum near \ion{Na}{1} and \ion{K}{1} before subtraction from the primary science target.}
\label{fig:samplespec}
\end{figure}

The raw echelle spectra from the HET were reduced using standard IRAF processes to remove the bias level of the CCD, flatten the field, remove scattered light, extract the spectra from the apertures, and determine the wavelength calibration from the ThAr lamp exposure.  The primary science targets, telluric observations, and ThAr exposures were all extracted to one-dimensional spectra.  We show examples of reduced but pre-processed primary science and telluric standard spectra in Fig.~\ref{fig:samplespec}.  The examples shown are for HD 189733 and its telluric standard HD 196867; the entire echelle orders containing \ion{Na}{1}, \ion{Ca}{1}, and \ion{K}{1} are shown.

Before subtracting the telluric standard observations from the primary science observations, we must ``clean" them of weak stellar and interstellar lines.  We fit simple models to the extracted telluric star observations, simultaneously modeling the star and the telluric absorption lines using IDL programs.  These programs used the MPFIT architecture \citep{Markwardt2009} for $\chi^2$ minimization.  The star is modeled with a base stellar model \citep{Koesterke2008, FM2005, CastelliKurucz2006} with free parameters of a velocity shift and rotational parameter $v\sin{i}$.  Absorption from the Earth's atmosphere is modeled with a base telluric model with free parameters of a velocity shift and optical depth.  We examine the residuals of these fits for additional absorption.  Among our three telluric stars, we find various combinations of interstellar absorption and additional stellar absorption (likely from a companion to the telluric standard star), with wavelengths corresponding to the Na D doublet and \ion{K}{1} $\lambda7698$.  These lines are fit for column density and $b$-value (i.e., Doppler parameter) in order to characterize their profile imprint.

For a given telluric star, all observations are fit individually at first, with all free parameters described in the previous paragraph allowed to vary.  Then the median fits of $v\sin{i}$ for the star and the median column density and $b$-value combination of the additional interstellar and stellar companion components are selected.  These values are then fixed and the observations are re-fit for velocity shifts and telluric optical depth.  This new, standardized best fit of the stellar model and interstellar/stellar companion absorption are then divided from the telluric observation.  The telluric absorption is modeled in order to aid the fit but is not applied to this division, resulting in a telluric observation ``cleaned" of stellar and ISM absorption imprints.

These ``cleaned" telluric observations are then removed from the primary science observations using standard IRAF procedures.  In cases where our custom telluric star was not observed in a given night, we compare the primary observation to all available telluric observations of the same custom telluric star over a region dominated by telluric lines and select the match with the lowest $\chi^2$ (after a fit for velocity shift and relative flux scale).

Removing the telluric lines removes the majority of the imprint of the blaze function on the primary science observations.  We then use a low-order polynomial to complete the normalization and fix the scale to unity.  Flux errors are calculated based on photon and read noise, and scaled by the normalization factors that are introduced from the telluric subtraction and additional normalization.

We coadd all observations into two sets, in-transit and out-of-transit.  Based on the ephemeris we use (see Table \ref{table:systemparams}), some observations cross through first or fourth contact during the exposure.  We classify these observations as either in- or out-of-transit based on the midpoint of each observation.  Observations with a midpoint during ingress or egress are classified as ``in-transit," although we make note of how many observations this applies to in Table \ref{table:observations}.  We also test whether or not using only observations with a midpoint during full transit (i.e., between second and third contact) makes a difference in the final results (see \S\ref{s:results}).

During our coadd, we apply two higher-order corrections.  First, we observe that the resolution of the HET at our settings varies with time and with wavelength.  This variation is most severe at short wavelengths of a given order, with the greatest variation before December 2007.  We empirically measure the resolution by fitting the width of unsaturated lines in the ThAr lamp exposure.  We measure lines over the entire wavelength region of each order and then fit the resolution as a function of wavelength in each order for each night with a low-order polynomial.  At each resonance line of interest, we find the minimum resolution over all observations and degrade the remaining observations to a chosen resolution slightly less than this using a Gaussian kernel (e.g., we find the resolution of all observations to be at least slightly greater than $R=65,000$ near Na D, and thus we degrade all observations to $R=65,000$).

The second correction that we apply is a wavelength correction between the various observations.  This corrects for both any overall shift in the observations and higher-order variations in the wavelength solution that can be observed by inspecting the stellar lines.  We perform a cross-correlation on strong stellar lines and fit a spline through the cross-correlation values.  This correction is applied to the wavelength arrays and the flux and error arrays are interpolated onto the new wavelength grids.  The final coadding is weighted by the errors at each point in the spectrum, with the errors based on photon statistics and read noise, as described above.

\section{ANALYSIS}
\label{s:analysis}

\subsection{Transmission Spectra}
\label{ss:differenced}
As described in the previous section, we make two coadds---one of all in-transit spectra, and one of all out-of-transit spectra.  We perform another wavelength correction as described in \S\ref{ss:reduction} in order to ensure that the two wavelength arrays are on identical grids and correct for any wavelength solution differences that still exist between the two sets of coadded spectra.  We then divide the in-transit spectrum by the out-of-transit spectrum and subtract one (to shift from a spectrum normalized to unity to one normalized to zero) to calculate our ``differenced" or ``transmission" spectrum.

We averaged the transmission spectra over several different fixed wavelength regions or ``wavebands"---a central waveband centered around the line we want to analyze and two additional wavebands of equal size, immediately to the blue and red of the central waveband.  The absorption lines and the wavebands widths are given in Table \ref{table:absorptionlines}.  The center of the waveband is determined from the minimum point of the absorption line, which varies slightly between our targets and when examining subsets of the data (see \S\ref{ss:montecarlo}); in the case of the \ion{Na}{1} doublet, the midpoint of the waveband is the midpoint between the minima of the two doublet lines.  The total absorption level is then the central waveband minus the average of the blue and red wavebands.  We carry through, using standard error propagation, the formal statistical errors on each point through the differencing and averaging steps to calculate a statistical error of the total absorption level.

\subsection{Estimating Errors with Monte Carlo Methods}
\label{ss:montecarlo}
Our final results are sensitive to any systematic errors that may be present in the reduction process---extraction of the spectra, cleaning of the telluric standard spectra, removing the telluric imprint from the primary science spectra, coadding the spectra, and differencing the coadded in-transit vs.~out-of-transit spectra.  \citet{Redfield2008} suspected that these possible systematic errors are larger than the formally propagated statistical errors.  Thus, \citet{Redfield2008} devised an empirical check on systematic errors.  They referred to their method as an ``Empirical Monte Carlo" method (hereafter EMC), and we repeat this method of analysis here.

We use three versions of the EMC analysis.  In each version, we perform the analysis described in \S\ref{ss:differenced} but instead of the total coadded in-transit spectrum, a subset of observations is taken, coadded, and a differenced spectrum is calculated relative to a single ``master" spectrum.  We perform thousands of iterations of this process for each EMC method in order to examine the centroid and width of the distribution and see how this compares with the master transmission spectrum of each target and absorption line combination.

The first EMC method we refer to as the ``in-out" method, where we randomly select a subset of in-transit spectra, coadd the selections using the same methods as above, and then calculate a transmission spectrum relative to the total coadded out-of-transit spectrum.  The number of selected spectra in the in-transit subset is randomly varied, from five up to one-half of the total number of in-transit spectra available, e.g.,~HD 189733 has 35 in-transit and 172 out-of-transit observations (observations which are not discarded; see Table \ref{table:observations}), so between 5 and 18 in-transit spectra are chosen for the subset.  We expect the in-out method to result in a distribution with a centroid consistent with the absorption level calculated including all observations.

The second method is the ``out-out" method, where the calculated transmission spectrum is not a true transmission spectrum, but a comparison of a randomly selected subset of out-of-transit observations, compared to the complete coadded out-of-transit spectrum.  The number of observations in the subset is the same as the number of total in-transit observations (e.g.~35 for HD 189733).  In principle this method should show a normal distribution centered at 0 (no absorption) with a width that indicates a robust measurement of systematic errors in our data processing and the system itself (e.g.~stellar variability).

The final method is the ``in-in" method.  As with the out-out method, a random subset of in-transit observations are compared to the complete in-transit spectrum.  The number in the subset is chosen such that the number of subset observations to total number in-transit observations is the same as the ratio of the number in-transit observations to the number of out-of-transit observations (e.g., $7/35\approx35/172$ for HD 189733).  As with the out-out method, the in-in method should be centered at no absorption.  The width of the distribution, however, is less meaningful because of the lower signal-to-noise level resulting from the smaller number of observations.

This trifecta of methods is different from the more common ``bootstrap" or ``quick-and-dirty" Monte Carlo \citep{Press1992}.  In the bootstrap Monte Carlo, a dataset of an equal number of points as the original set is chosen, with replacement possible.  ``Replacement" means that a randomly chosen point (or in our case, observation) is ``put back (replaced) into the bin" from which the points are drawn, and therefore duplication is possible, and in a typical iteration $1/e\approx37\%$ points in the dataset will be duplicated.  The methods described above use randomly chosen subsets without replacement, and therefore bear some similarities to a ``jackknife" method \citep{Wall2003}.  There are two primary reasons that we use our ``out-out" EMC method to determine the error.  The first is that by using the ``out-out" method we are explicitly removing the effects of planetary transit when we attempt to quantify the astrophysical noise (in addition to the statistical errors and reduction systematics).  The second reason is that by using the ``out-out" method, we are able to draw upon a larger set of points for the Monte Carlo analysis.  However, we also test the more traditional bootstrap Monte Carlo, in a way that is most analogous to our ``in-out" method.  Using HD 189733b as an example, the boostrap method would entail a random sampling of 35 ``in-transit" observations, with replacement (i.e., duplication) allowed, and a sample of 172 ``out-of-transit" observations, again with replacement.  We discuss these tests along with the rest of the results in \S\ref{s:results}.

\subsection{Confirming Detections with a Control Line}
\label{ss:controlline}
\citet{Redfield2008} also tested the potential objection that there may be systematic errors simply from the process of coadding and differencing spectra surrounding strong stellar absorption lines.  In order to test this, they calculated a transmission spectrum at \ion{Ca}{1} $\lambda6122$, which is a reasonably strong stellar line in our target systems but not a resonance line that is expected to manifest in hot Jupiter-type atmospheres according to models (it is expected to be condensed out, e.g., into CaTiO$_3$).  We also calculate a transmission spectrum of this line for all four targets.

The wavelength region near this line is utterly devoid of telluric features, so our data analysis process is modified slightly in that we do not perform telluric subtraction.  This is advantageous in that suboptimal telluric subtraction will not influence the results.  It is disadvantageous in that it means the control line analysis is different from the Na and K line analysis in this particular step (all other steps are identical).  However, it provides a consistency check on the quality of our telluric subtraction.

In addition to calculating a total transmission spectrum, we also run the EMC analysis for this line.  We expect a null master measurement (no absorption) and an EMC distribution centered around zero as well to confirm that our measurements are not simply a typical systematic byproduct of differencing around strong stellar lines.

\section{RESULTS}
\label{s:results}
\begin{figure}[t!]
\begin{center}
\epsscale{1.00}
\plotone{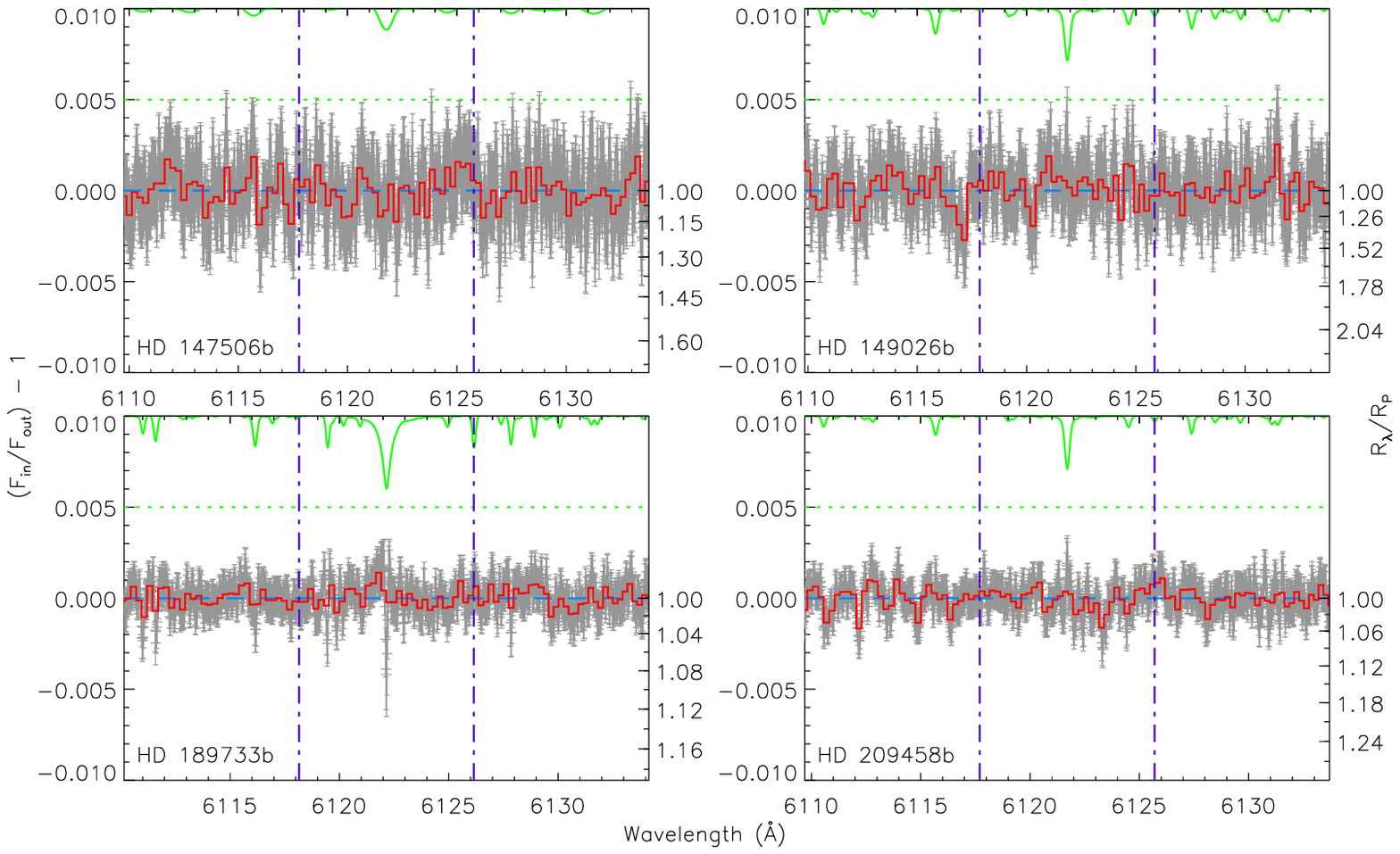}
\end{center}
\caption[]{Same as Fig.~\ref{fig:kidiffspec} but for \ion{Ca}{1} $\lambda6122$.}
\label{fig:caidiffspec}
\end{figure}

We have four targets that we examine for evidence of \ion{Na}{1} ($\lambda\lambda$5889,5895), \ion{Ca}{1} ($\lambda$6122), and \ion{K}{1} ($\lambda$7698).  We expect \ion{Na}{1} and \ion{K}{1} to exist in atomic form in hot Jupiter atmospheres and for these ground-state resonance lines to exhibit absorption.  However, as described in the previous section (\S\ref{ss:controlline}), this is not true of \ion{Ca}{1}, which functions as a control line.  The transmission spectra for each line and target are shown in Fig.~\ref{fig:naidiffspec}--\ref{fig:kidiffspec}.  These are shown in order of increasing wavelength:  \ion{Na}{1}, \ion{Ca}{1}, and \ion{K}{1}, respectively.  Our measurements are shown in Table \ref{table:results}.

\begin{figure}[t!]
\begin{center}
\epsscale{1.00}
\plotone{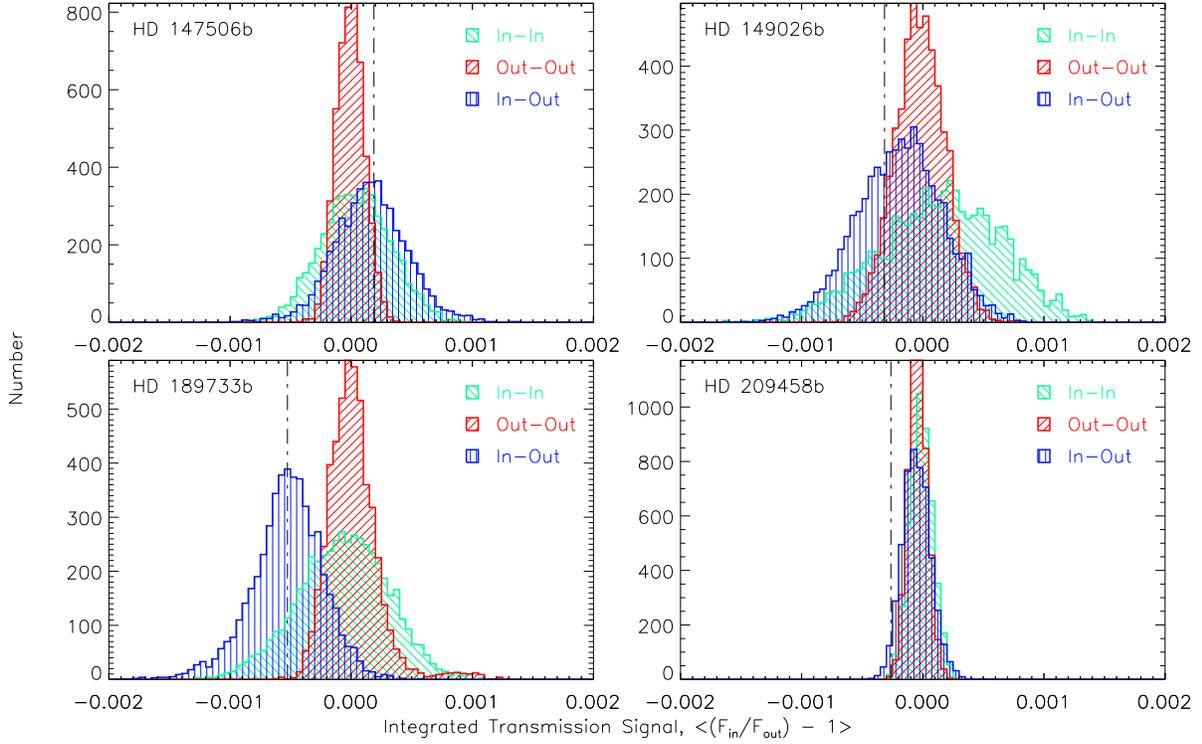}
\end{center}
\caption[]{EMC analysis results for \ion{Na}{1} $\lambda5889, 5895$.  Clockwise from upper left:---HD 147506, HD 149026, HD 209458, HD 189733.  The distributions for the three versions of the EMC (see \S\ref{ss:montecarlo}) are plotted in different colors, shown in the key:  green is the ``in-in" distribution, red is the ``out-out" distribution, and blue is the ``in-out" distribution.  5000 iterations are used in each case.  The master ``in-out" measurement is shown as a vertical dot-dashed line.  Note that the scale of the x-axis is chosen to be consistent between this plot and the similar plots in Fig.~\ref{fig:caiemc} and Fig.~\ref{fig:kiemc}, because the width of the distribution is the important quantity.  Conversely, the scale of the y-axis is varied in each panel to show maximum detail.}
\label{fig:naiemc}
\end{figure}

\begin{figure}[t!]
\begin{center}
\epsscale{1.00}
\plotone{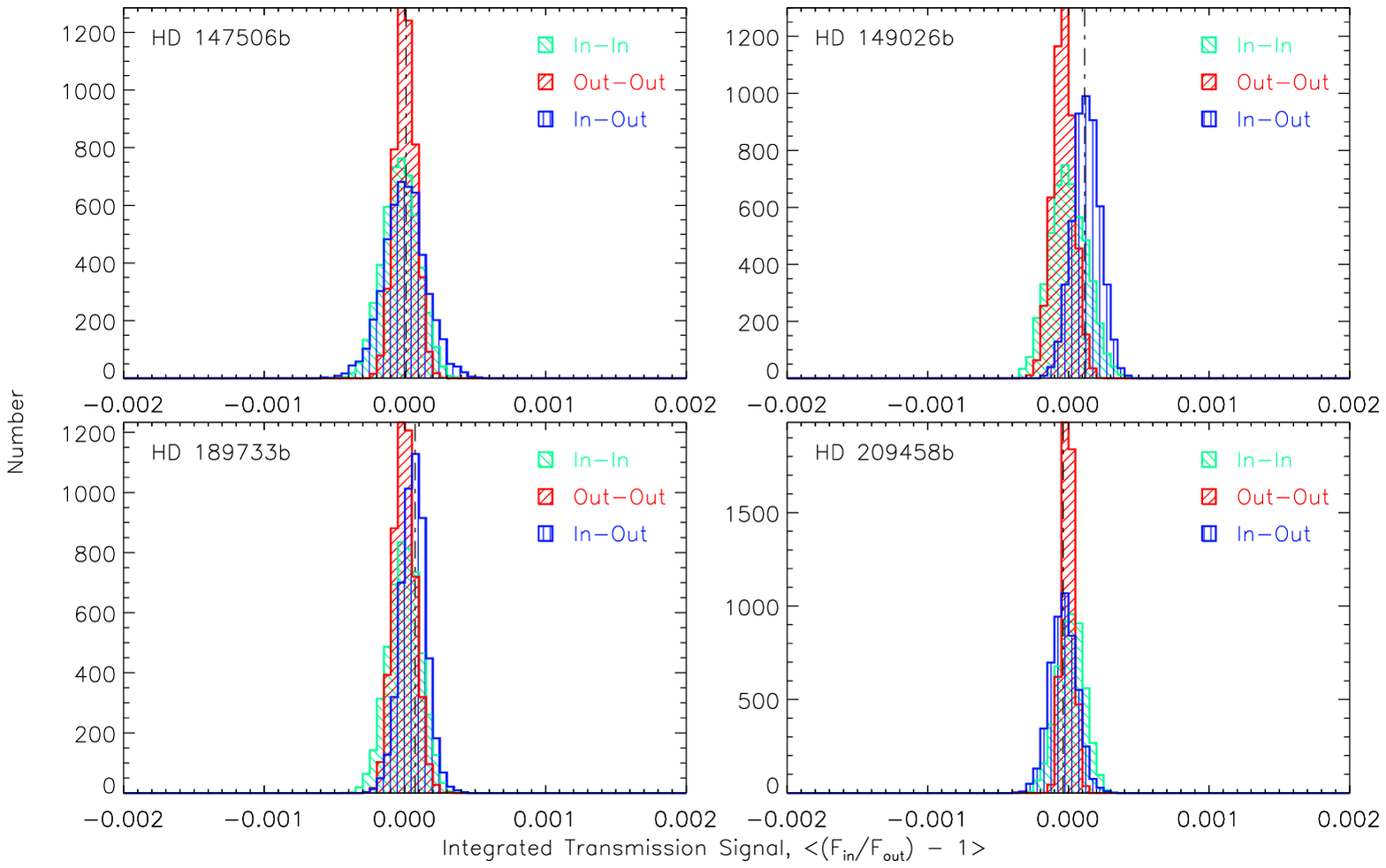}
\end{center}
\caption[]{Same as Fig.~\ref{fig:kiemc} but for \ion{Ca}{1} $\lambda6122$.}
\label{fig:caiemc}
\end{figure}

\begin{figure}[t!]
\begin{center}
\epsscale{1.00}
\plotone{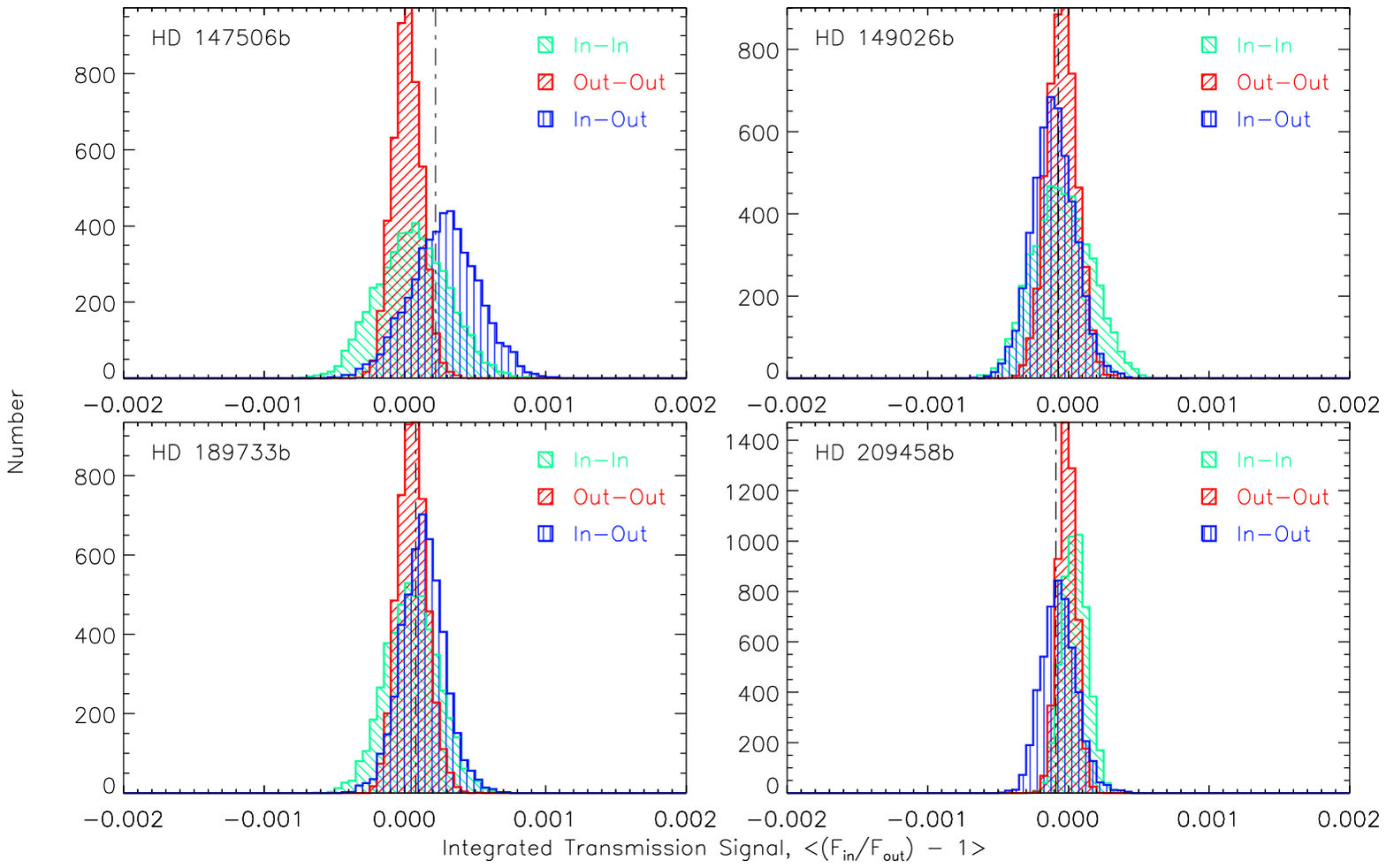}
\end{center}
\caption[]{Same as Fig.~\ref{fig:kiemc} but for \ion{K}{1} $\lambda7698$.}
\label{fig:kiemc}
\end{figure}

The corresponding EMC error analyses are shown, in the same order, in Fig.~\ref{fig:naiemc}--\ref{fig:kiemc}.  In the discussion of the subsequent sections, we will use the width of the ``out-out" EMC distribution as the 1$\sigma$ error.  We first discuss our results for \ion{Ca}{1}, the control line.

\subsection{\ion{Ca}{1} Control Line}
\label{ss:cairesults}
As discussed in \S\ref{ss:controlline}, we do not expect absorption due to \ion{Ca}{1}, as calcium should be condensed out at the atmospheric temperatures of our targets.  In three of the four targets, the master measurement of flux at the control line is consistent with zero to within the 1$\sigma$ error, as we expect.  In HD 149026b, the master measurement deviates from zero, at a significance level of 1.48$\sigma$.  The deviation is positive, indicating ``emission" as opposed to absorption, but such a result, if robust, is not physically meaningful for transits.  Minor features can be seen in the top right panel of Fig.~\ref{fig:caidiffspec}, consistent with the calculated significance level.

Using a standard bootstrap Monte Carlo method provides similar results---3 of the 4 targets are within 1$\sigma$ of zero, and HD 149026b is within 2$\sigma$.  Notably, the error (i.e., the distribution width) produced by the bootstrap method is slightly larger than the ``out-out" EMC width, but we have explained in \S\ref{ss:montecarlo} our reasons for adopting the latter as our error.  In addition, if we perform a master transmission measurement using only observations with a midpoint between second and third contact, the results are again essentially the same.  The results for all four targets taken as a whole confirm that statistically significant absorption and emission profiles are not inevitable, or even probable, simply as a byproduct of our reduction and analysis processes.  This provides confidence in the results that we do detect in other lines.


\subsection{\ion{Na}{1}}
\label{ss:nairesults}
\begin{figure}[t!]
\begin{center}
\epsscale{1.00}
\plotone{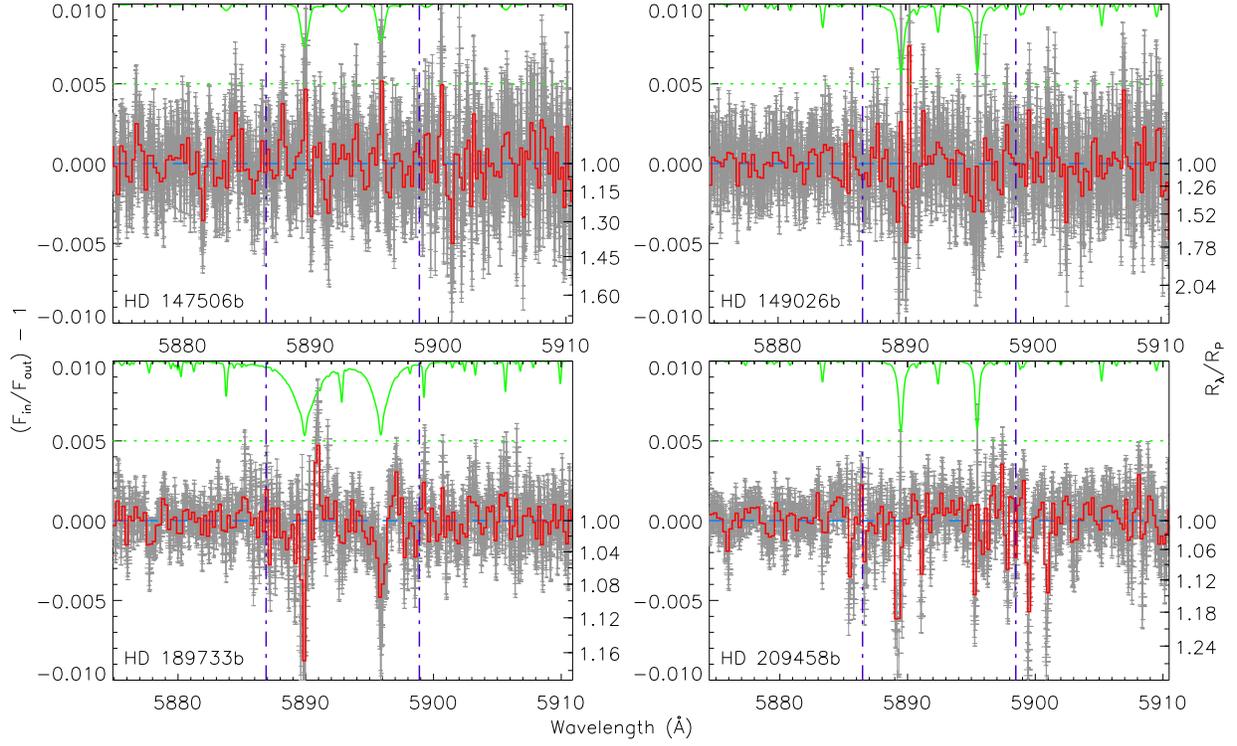}
\end{center}
\caption[]{The various transmission spectra at \ion{Na}{1} $\lambda5889, 5895$.  Clockwise from upper left:---HD 147506, HD 149026, HD 209458, HD 189733.  The binned transmission spectrum is shown in red, with the error bars of the unbinned spectrum shown in gray beneath the binned spectrum.  For reference, the master out-of-transit spectrum of the star is shown at the top of each panel in green, scaled down to fit the plot, with a dotted green line showing the zero level and the top of each plot being unity in the normalized spectrum.  A blue dashed line shows the zero point for the transmission spectrum.  The purple vertical dot-dashed lines define the bandpass that is integrated to make our absorption measurements; from these lines to the edges of the plot define the comparison bandpasses.  The right-hand y-axis shows the absorption in terms of relative apparent planetary radii.  Values less than 1.0 are not shown on this axis because they are unphysical (and should only occur due to noise or artifacts).}
\label{fig:naidiffspec}
\end{figure}

We detect absorption at \ion{Na}{1} in HD 189733b at a significance of 3.0$\sigma$, with an absorption level of $(-5.26\pm1.69)\times10^{-4}$.  This is consistent to within 0.55$\sigma$ with the $(-6.72\pm2.07)\times10^{-4}$ absorption found by R2008 and confirmed by recent HST measurements (D.~Sing, 2010, private communication; Huitson, et al., submitted).  The profiles can be seen in the bottom left panel of Fig.~\ref{fig:naidiffspec} and are comparable to the profiles seen in R2008.  We note that our original data is the same as that used by R2008.  However, our data reduction and analysis were not identical, using a quasi-independently derived set of IRAF reduction scripts and IDL analysis programs in order to apply these programs to the broader dataset of additional targets and \ion{K}{1} analysis.

HD 209458b is also known to show \ion{Na}{1} absorption \citep{Charbonneau2002, Snellen2008}.  The \citet{Charbonneau2002} detection was $(-2.32\pm0.57)\times10^{-4}$ in an absorption band identical to ours, while \citet{Snellen2008} used a variety of bandpasses that are not directly comparable to our results.  We find absorption at a level of $(-2.63\pm0.81)\times10^{-4}$, nearly identical to the \citet{Charbonneau2002} measurement.  However, the transmission spectrum (bottom right panel of Fig.~\ref{fig:naidiffspec}) is less clear, with only the narrowest features seen at the \ion{Na}{1} lines, along with several other narrow features in the red comparison band.  In addition, of the 12 EMC analyses (4 targets and 3 absorption lines), this is the only case where the centroid of the ``in-out" distribution is discrepant with the master measurement, as the ``in-out" centroid is consistent with zero ($(-0.57\pm0.81)\times10^{-4}$).  A possible explanation is the narrow features in the red comparison band, as our master measurement is based on the integration of the central bands defined in Table \ref{table:absorptionlines} minus the average of equivalent-width bands to immediately to the blue and red.  If we modify the calculation so that we only subtract the blue comparison band for both the master measurement and the EMC results, then the centroid of the ``in-out" distribution changes to $-1.44\times10^{-4}$ and the master measurement to $-3.30\times10^{-4}$.  These results are 0.9$\sigma$ and 1.0$\sigma$ discrepant from the \citet{Charbonneau2002} results, respectively, but still discrepant from each other.  It is possible that the narrow features in the red comparison band are artifacts of the telluric subtraction process, but this is not clear.  The three strongest of these features roughly match known telluric lines, but there are inconsistencies in the velocity offsets.  Furthermore, we do not see features in the transmission spectrum at the wavelength of other telluric lines of nearly comparable strength.  We have also examined the individual spectra and are unable to determine if and in what spectrum the possible subpar telluric subtraction occurs.  Therefore, it is unclear if poor telluric subtraction could be responsible for these features.  Currently we do not have a clear explanation for the cause of these features or the internal inconsistency between the master measurement and the EMC centroid.

Our measurement for HD 147506b indicates a 1.6$\sigma$ surplus.  However, no feature is apparent in the transmission spectrum (top left panel of Fig.~\ref{fig:naidiffspec}).  Conversely, HD 149026b seems to show absorption at a comparable level of 1.5$\sigma$.  Though the significance levels are similar, the latter has a greater magnitude and a larger error.  Additionally, absorption features are seen at both \ion{Na}{1} lines (top right panel of Fig.~\ref{fig:naidiffspec}).  We hypothesize that this absorption may be real, though additional observations would be required to confirm this.

Similar to the results for \ion{Ca}{1}, the bootstrap Monte Carlo results for \ion{Na}{1} are consistent with our master and EMC results.  In the case of HD 209458b, the bootstrap distribution is like our ``in-out" EMC distribution in that it is consistent with zero and inconsistent with the master measurement.  The bootstrap errors are also slightly larger ($<50\%$ on average) than the ``out-out" EMC results.

If ingress and egress observations are discarded, the results are still consistent with our original master measurements to within $<$1$\sigma$.  We do note that the measured HD 189733b \ion{Na}{1} absorption increases to $-6.50\times10^{-4}$.  This is notable in part because of our four targets HD 189733b has the greatest percentage of ingress/egress observations that we must discard (due to the planet's larger R$_{\rm P}$/R$_{\rm *}$ and shorter orbital period, the ratio of time spent in ingress/egress to full transit is larger).  Thus, if signal dilution from the ingress/egress observations is a problem for our original master measurement, we would expect the effect to be strongest in this case.  The \ion{Na}{1} absorption levels for HD 149026b and HD 209458b, however, decrease to $-2.04\times10^{-4}$ and $-1.87\times10^{-4}$, respectively.

\subsection{\ion{K}{1}}
\label{ss:kiresults}
\begin{figure}[t!]
\begin{center}
\epsscale{1.00}
\plotone{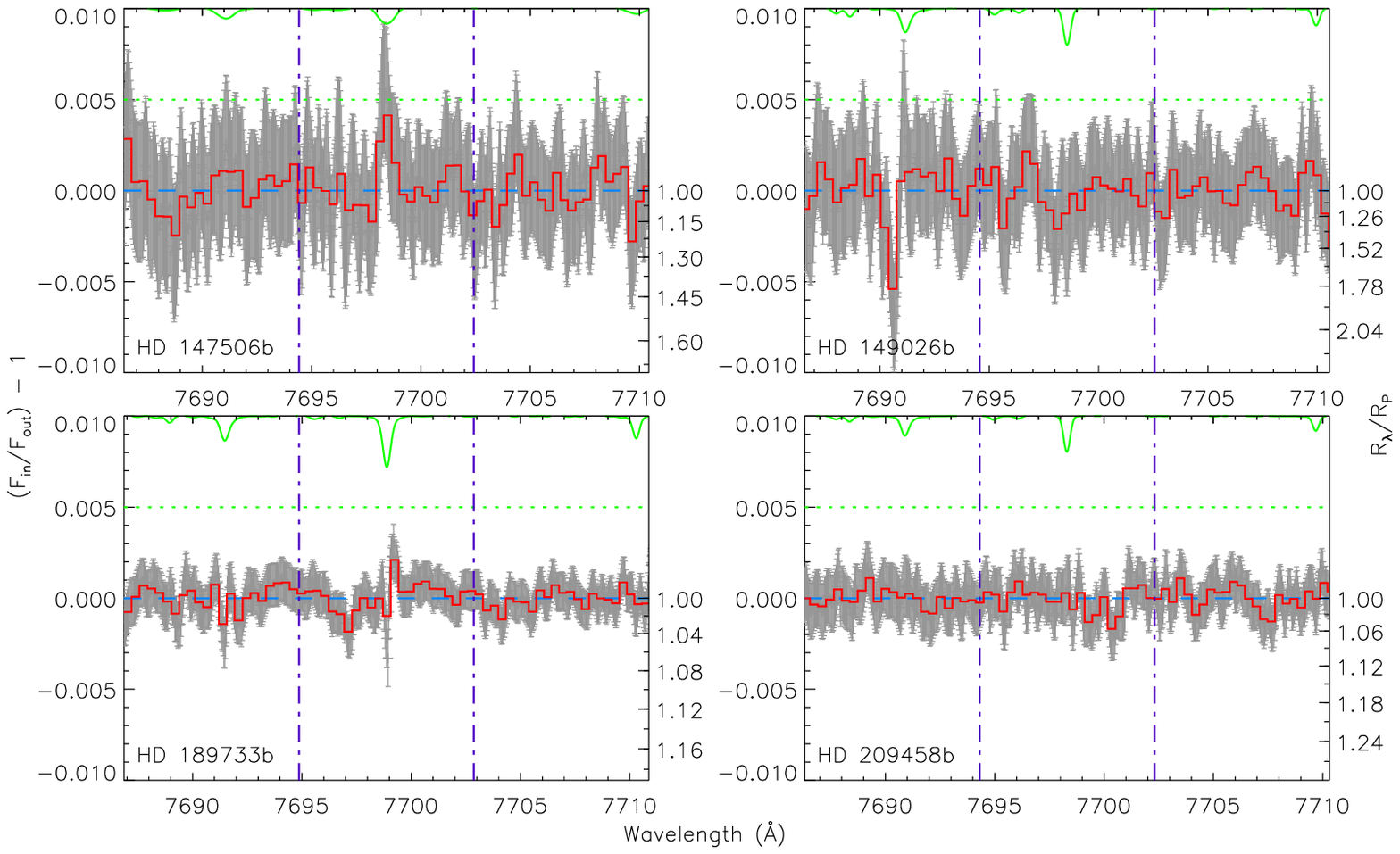}
\end{center}
\caption[]{Same as Fig.~\ref{fig:kidiffspec} but for \ion{K}{1} $\lambda7698$.}
\label{fig:kidiffspec}
\end{figure}

It has only been in the last year that \ion{K}{1} has been detected in exoplanetary atmospheres \citep{Colon2010,Sing2011}.  Examining Fig.~\ref{fig:kidiffspec}, we do not see any cases of clear \ion{K}{1} absorption, though our measurement for HD 209458b is $(-0.89\pm0.67)\times10^{-4}$, which is $1.3\sigma$ significant.  HD 189733b shows no feature and is consistent with zero to within the 1$\sigma$ errors.

The other two targets, however, do show features worth mentioning.  HD 149026b show a feature in the blue comparison band that is visually significant relative to the noise.  This feature is not centered on any feature in the stellar spectrum, and is too far from \ion{K}{1} in velocity space to be blueshifted \ion{K}{1} absorption.  If we neglect the blue band and compare the central measurement to only the red band, then the master measurement of the \ion{K}{1} is $0.02\times10^{-4}$, miniscule compared to the $1.09\times10^{-4}$ error.  The feature itself, integrated over the entire blue band, is also less significant than the 1$\sigma$ error.

HD 147506b is the most difficult to understand case.  There is an emission-like feature centered on \ion{K}{1} in the transmission spectrum (upper left panel, Fig.~\ref{fig:kidiffspec}).  According to our EMC calculations, this feature is 2.1$\sigma$ significant.  There is no easily understood physical reason why a transmission spectrum should show apparent emission.  However, we note that of our four targets, HD 147506b has relatively low overall S/N in the raw observations (comparable to HD 149026b but much lower than HD 189733b and HD 209458b).  Its stellar features are also the broadest and shallowest of our entire sample \citep[due to its $v \sin{i}$ of $20.8\pm0.3\kmpers$, $\geq3\times$ the other three stars in our sample;][]{Pal2010, Butler2006}, which complicates the line cross-correlation required for the wavelength corrections we use to match up the master in- and out-of-transit spectra before performing the differencing step.  Furthermore, the echelle order containing \ion{K}{1} for HD 147506b was by far the most uncertain dataset qualitatively, as we discarded multiple observations due to reduction artifacts (spikes and dips in the nearby spectrum) of unknown origin.  Though these were discarded, it is unclear if any observations remain that unduly influence the measurement here.  For these reasons, and the fact that feature is only 2.1$\sigma$ significant, we argue that this feature does not correspond to any real physical effect, and is most likely isolated to the analysis of this particular absorption line in this single target.

Once again, considering the bootstrap Monte Carlo or only the transits only between second and third contact do not appreciably change any of the results.  We note, however, that the much larger error from the bootstrap method would make the HD 147506b ``emission" only 1.3$\sigma$.

\section{DISCUSSION}
\label{s:discussion}
\begin{figure}[t!]
\begin{center}
\epsscale{1.00}
\plotone{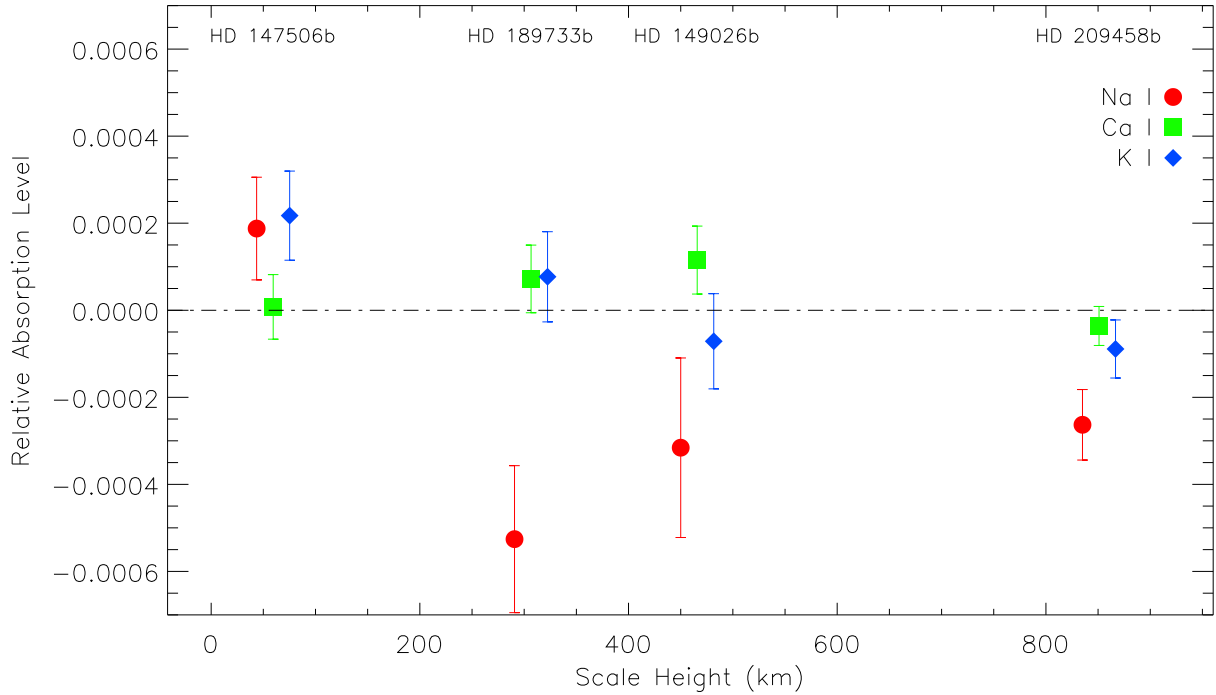}
\end{center}
\caption[]{A plot of all absorption results as a function of scale height of the atmosphere, with different color-coded symbols for each element.  The scale height is calculated using Jupiter's mean molecular weight of 2.22 g/mol and the radiative equilibrium temperature of each planet.  \ion{Na}{1} is plotted at the proper scale height for each planet, while \ion{Ca}{1} and \ion{K}{1} are slightly offset for clarity.  1$\sigma$ error bars are shown.  A dot-dashed line is also shown at 0 (no absorption) for reference.}
\label{fig:scaleheight}
\end{figure}

\begin{figure}[t!]
\begin{center}
\epsscale{1.00}
\plotone{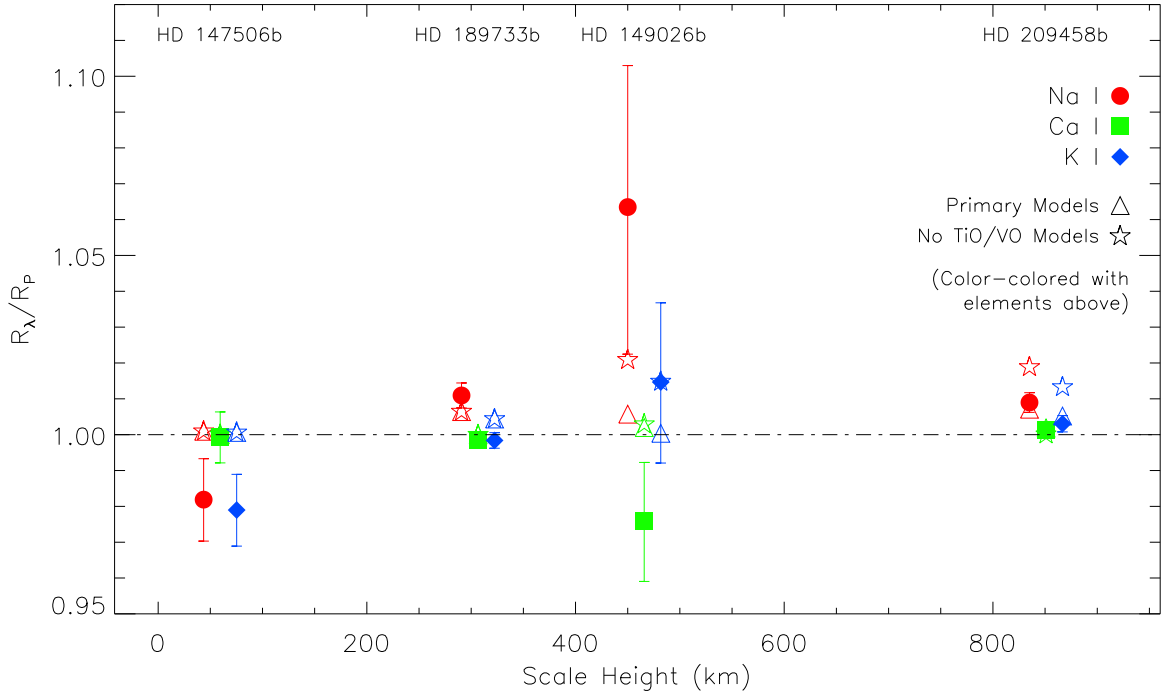}
\end{center}
\caption[]{Same as Fig.~\ref{fig:scaleheight} except with the transmission signal measurements converted into apparent planetary radius, relative to each planet.  Note that the $\lambda$ in ${\rm R}_{\lambda}$ is not a specific wavelength in this case but an integrated value based on the bandpasses defined in Table \ref{table:absorptionlines}.  A key to the symbols is shown on the plot; solid points are measured values, while open points are the model values discussed in the text, color-coded and aligned on the x-axis with the measurement it is intend to be compared to.}
\label{fig:scaleheightrp}
\end{figure}

We synthesize our results for all three elements in Fig.~\ref{fig:scaleheight}.  In this plot, it is easy to see the consistency of \ion{Ca}{1} with zero, which we expect as our control line.  The other results discussed in \S\ref{ss:nairesults}--\ref{ss:kiresults} can also be seen, including the $\geq3\sigma$ detections of \ion{Na}{1} in HD 189733b and HD 209458b and the very tentative ($1.5\sigma$) indication of \ion{Na}{1} absorption in HD 149026b.  Fig.~\ref{fig:scaleheightrp} shows the same data from a different perspective; it is similar to Fig.~\ref{fig:scaleheight}, except that plotted on the y-axis is the apparent radius of the planet, averaged over the bandpasses defined in Table \ref{table:absorptionlines}.  Fig.~\ref{fig:scaleheightrp} also includes integrated model values, which we discuss later in \S\ref{ss:models}.

\subsection{Dependence on Scale Height}
\label{ss:scaleheight}
Scale height is used in Fig.~\ref{fig:scaleheight} and \ref{fig:scaleheightrp} because it is a straightforward measurement that should correlate with the amount of observed absorption, all other factors being equal.  It is obviously difficult to pull out any clear trends from these two plots given that only two points are statistically different than zero ($\geq3\sigma$).  However, we can say that our detections (\ion{Na}{1} in HD 189733b and HD 209458b) and our nominal detection (\ion{Na}{1} in HD 149026b) are consistent with what we expect from consideration of atmospheric scale height.  We observe nothing for HD 147506b, which is the most massive of our four planets (but similar in size to HD 189733b and HD 209458b), thus giving it the largest surface gravity and smallest scale height, while the other three planets, with much larger scale heights, show at least the hint of absorption.

There are other properties beyond scale height that determine the detectability of atmospheric absorption.  One important property is the metal abundance.  HD 149026 has the highest metallicity, $\rm{[Fe/H]}=0.36\pm0.08$, of all the central stars of our targets.  HD 149026b can likewise be inferred to be particularly metal rich based on bulk density arguments \citep{Sato2005}.  While there is no guarantee that these quantities would translate into an enhanced metal abundances in the atmosphere, bulk density and atmospheric abundances are correlated in the Solar System \citep{Lodders}.  It is therefore possible that HD 149026b might have an enhanced atmospheric \ion{Na}{1} abundance, which could be responsible for the large (albeit statistically tentative) ${\rm R_{\lambda}/R_P}$ that we derive for the 12 \AA{} wide band around \ion{Na}{1} in this target.

\subsection{Temperature Inversions}
\label{ss:inversions}
Another important consideration is the issue of atmospheric temperature inversions.  Spitzer observations of many extrasolar planets have revealed two classes of planets:  those whose IR spectra are fit well by models with H$_2$O and CO in absorption, and planets whose spectra are better fit by models with those same molecules in emission, indicating a temperature inversion \citep{Hubeny2003}.  HD 189733b is in the former category \citep{Charbonneau2008} and HD 209458b is in the latter class \citep{Knutson2008}.  There has been comparatively little IR work on HD 147506b and HD 149026b, with no Spitzer observations of the former \citep{SeagerDeming2010} and only 8 $\mu$m work on the latter \citep{Harrington2007, Knutson2009}.  Consequently, \citet{Knutson2010} classifies these planets as unknown with respect to temperature inversions.

A temperature inversion is relevant to \ion{Na}{1} and \ion{K}{1} absorption because it requires an additional absorber (e.g.~VO/TiO or clouds) that may obscure the absorption of other major opacity sources.  \citet{Fortney2003}, \citet{Burrows2007}, and \citet{Desert2008} have invoked such an absorber as a possible reason why the absorption in \ion{Na}{1} in HD 209458b is less than expected \citep{Charbonneau2002} and less than that of HD 189733b \citep{Redfield2008}, which does not appear to have an inversion.  If this is the case, then our possible detection of \ion{Na}{1} in HD 149026b---with a scale height greater than HD 189733b but much less than HD 209458b---might nominally predict that it does not have an inversion.  \citet{Knutson2010} has proposed that increased stellar UV flux may destroy whatever unknown material is responsible for the inversion.  Notably, the central star HD 149026 has a smaller UV flux more consistent with the inversion cases shown in \citet{Knutson2010}.  However, \citet{SeagerDeming2010} note that there are counterexamples of high stellar activity and assumed inversions, most notably XO-1b.  They also note that many cases of inferred temperature inversions have adequate alternative explanations to within the uncertainties, e.g., an increased CH$_4$ abundance relative to CO, and that three-dimensional circulation models are inadequate to explain all of the Spitzer/IRAC data, even with inversions included \citep[see][and references therein]{SeagerDeming2010}.  We also note that other factors likely play a role in depleting the observed atmospheric abundances, including ionization and condensation.  Both \citet{Sing2008b} and \citet{VidalMadjar2011} conclude that there must be either an ionization or condensation effect in HD 209458b; in particular, the latter invokes both explanations at different layers in the atmosphere.

\subsection{Exospheric Absorption}
\label{ss:exospheres}
We have assumed that any detection of \ion{Na}{1} or \ion{K}{1} would come from the bound atmosphere of the planet.  Measuring wavelength-dependent apparent planetary radius (i.e.,~${\rm R_{\lambda}}$) larger than the Roche radius would indicate that some absorption comes from the unbound exosphere, e.g.~the hydrogen detection in HD 209458b made by \citet{VidalMadjar2003, VidalMadjar2004} through Ly$\alpha$ measurements.  However, the converse is not true; if ${\rm R_{\lambda}}$ lies inside the Roche radius, the absorption may be either bound or unbound depending on the details of the absorption (e.g.,~abundance, saturation, etc.).  Based on our transmission spectra in Fig.~\ref{fig:naidiffspec}--\ref{fig:kidiffspec} and averaged results in Fig.~\ref{fig:scaleheightrp}, we do not find evidence that any of the absorption is from the exosphere of the planet.  However, given that both HD 189733b and HD 209458b appear to have some escaping atmosphere, and are possibly in a state of ``blow-off" \citep[detected in Ly$\alpha$ by][respectively, for HD 189733b and HD 209458b]{LecavelierDesEtangs2010,VidalMadjar2003,VidalMadjar2004}, it is possible that some \ion{Na}{1} absorption occurs in the exosphere due to the effects of hydrodynamic drag.

\subsection{Velocity Shifts of Atmospheric Lines}
\label{ss:velocityshifts}
\citet{Redfield2008} detected a substantial apparent velocity shift of $-38\kmpers$ in the \ion{Na}{1} lines of HD 189733b, relative to the \ion{Na}{1} lines of the star.  The authors suggested that the shift, if astrophysical in origin, may have been a combination of planetary motion and winds from the hot dayside flowing to the nightside.  More recently, such shifts have been measured in HD 209458b by \citet{Snellen2010}, although at a much smaller level of $2\kmpers$.  In this analysis, we measure apparent shifts in the lines by fitting Gaussian profiles to the features we noted near \ion{Na}{1} in HD 149026b, HD 189733b, and HD 209458b (although only the latter two are detected at significant levels; see \S\ref{ss:nairesults}).  We find shifts of approximately $-4$, $-37$, and $-12\kmpers$, respectively.  In the case of HD 189733b, our result is identical to the result of \citet{Redfield2008} to within the precision of the spectral resolution, which is $\sim5\kmpers$ for the HRS at its $R\sim60,000$ setting.

There is a qualitative difference between HD 189733b and the other two targets, in that the feature we associate with \ion{Na}{1} absorption is highly asymmetric.  This asymmetry is seen both in the reduction of this paper and \citet{Redfield2008}.  Before performing the step of differencing the in- and out-of-transit spectra, we perform wavelength corrections by performing cross-correlations on the lines between the two spectra to create a wavelength correction array.  If this cross-correlation is imperfect, then there will be an asymmetric feature created, with a dip to one side of the stellar line center and a spike to the other side, which is what is seen in our data (with a dip to the blue and a spike to the red, though the integration yields net absorption).  Whether or not this is the source of the artifact we see, we note that fitting a Gaussian to an apparent absorption feature when this type of artifact is present will artificially force the fit of line center to the side with the dip.  In HD 189733b, this is to the blue, whereas HD 149026b and HD 209458b do not show the same type of asymmetric artifact.  If instead of fitting Gaussian profiles we assume that line center of the atmospheric feature is the local minimum point in the transmission spectrum, then we find shifts of approximately 0, 1, and $-11\kmpers$ to HD 149026b, HD 189733b, and HD 209458b, respectively.  The results for HD 149026b and HD 209458b are consistent with the Gaussian results to within the $\sim5\kmpers$ resolution element.  We conclude that it is the nature of the HD 189733b artifact, seen both in this present work and in \citet{Redfield2008}, that causes the large apparent shift when measured through Gaussian line fits.  This suggests that the present data may be inadequate to robustly make precise measurements of atmospheric velocity shifts, especially when an asymmetric artifact of the differencing step is seen.

\subsection{Comparison with Models}
\label{ss:models}
In Fig.~\ref{fig:scaleheightrp}, there are additional points that show the integrated points of planetary atmosphere models.  One model includes absorption by TiO and VO (which we call the ``primary" model), while the other does not (the ``no TiO/VO" model).  The primary models, by their inclusion of these molecules, are imposing an absorber in order to create an atmospheric temperature inversion, which has been observed for several planets \citep[e.g.][]{Knutson2008}.  We note, however, that there is ongoing debate about whether or not TiO/VO can create a temperature inversion in a planetary amotsphere \citep[see, e.g.,][who argue the negative case]{Spiegel2009}.

For HD 147506b, it can be seen that the model points are all very close to 1 in ${\rm R_{\lambda}/R_P}$, i.e., no additional absorption from the atmosphere.  We noted in \S\ref{s:results} that the integration of the transmission spectrum for this star gave an excess at \ion{Na}{1} and \ion{K}{1}, though less than 3$\sigma$ in each case.  To the extent that these results are consistent with zero (in terms of relative absorption; 1 in ${\rm R_{\lambda}/R_P}$), they are also consistent with both the primary and no TiO/VO models.  We note that HD 147506b was included in this study because of its bright central star rather than planetary properties that suggested potential absorption.\footnote{The targets for this survey were selected in 2007 (earlier in the case of HD 189733b) when far fewer transiting exoplanetary targets were known, and all four stars remain among the brightest known to have a transiting planet.}  The fact that we see no obvious absorption is consistent with its much smaller scale height compared to our other targets.

HD 149026b shows the most significant spread in ${\rm R_{\lambda}/R_P}$, because its planetary radius is the smallest of our sample.  The primary model values are much closer to zero, but consistent with our measurements within 1.5$\sigma$ for \ion{Na}{1}, 1.6$\sigma$ for \ion{Ca}{1}, and 0.65$\sigma$ for \ion{K}{1}.  The model without TiO and VO provides a slightly better fit to both \ion{Na}{1} (within 1.1$\sigma$ vs.~1.5$\sigma$) and \ion{K}{1} (within 0.01$\sigma$ vs.~0.65$\sigma$).  We have suggested a possible detection of \ion{Na}{1}, significant at the 1.5$\sigma$ level but with a profile that suggests real absorption, and suggest that further observations should be performed to confirm or rule out this measurement.  Our current measurement implies a case closer to the model without TiO/VO, though it does not strongly differentiate between the two models.

The primary and no TiO/VO models are virtually identical for HD 189733b, because the temperature of this planet precludes the existence of these molecules.  The models predict less \ion{Na}{1} and more \ion{K}{1} absorption than we see---1.3$\sigma$ and 2.7$\sigma$ differences.  Despite the overprediction of \ion{K}{1} relative to our measurement, the magnitude of our prediction implies it is unlikely we would be able to detect any absorption at significant levels.

HD 209458b shows a good match at \ion{Na}{1} (within 0.68$\sigma$) and \ion{Ca}{1} (within 0.41$\sigma$) and a fair match at \ion{K}{1} (within 0.94$\sigma$) for the primary model.  The model without TiO/VO does much worse, as it is 3.6$\sigma$ and 4.5$\sigma$ discrepant from our measurements of \ion{Na}{1} and \ion{K}{1}, respectively.  This is very consistent with the fact that HD 209458b seems to have a temperature inversion \citep{Knutson2008}, implying the need for an absorber such as TiO, VO, or something else---though as noted above, there is significant debate about which absorber(s) can properly account for inversions.

As we noted in \S\ref{s:intro}, we are just now entering a time when we can do this sort of comparative atmospheric study for exoplanets.  In our small sample here, we have noted two cases that superficially favor different models (HD 149026b and HD 209458).  This drives home the importance of obtaining more data and creating data-driven models.

\subsection{Future Work}
\label{ss:futurework}
The spectra that we have obtained are a very rich resource for additional studies.  We are currently searching for H$\alpha$ absorption in these four targets (Jensen et al.~2011b, in preparation).  As noted above, both HD 189733b and HD 209458b have had detections of Ly$\alpha$, apparently from the unbound portions of their atmosphere, or exosphere.  Using low-resolution ACS spectra, \citet{Ballester2007} also detected absorption in HD 209458b corresponding to the Balmer edge of \ion{H}{1}.\footnote{By the Balmer edge of \ion{H}{1} we mean the combination of the $n=2\rightarrow n=\infty$ transition edge at 3646 \AA{}, the associated bound-free absorption from the $n=2$ state at nearby but shorter wavelengths, and bound-bound absorption from $n=2 \rightarrow n \gg 2$ at nearby but longer wavelengths.}  This implicitly predicts that H$\alpha$ absorption should occur in HD 209458b's atmosphere, though past searches have turned up empty \citep{Winn2004}.  A detection of H$\alpha$ ($n=2\rightarrow3$) in any exoplanetary atmosphere would be a first.  Combined with an existing Ly$\alpha$ detection it would be a significant result, allowing for the calculation of a hydrogen excitation temperature and implications for the atmospheric physics.  The many spectra that we have obtained over many epochs also present an interesting opportunity for work on stellar variability, including exploring activity-related interactions between the stars and planets in these systems \citep[e.g., the preliminary work reported by][]{Arai2010}.  Most importantly, additional targets remain a possibility for this program.  Observations of one such target, a hot Neptune-class planet, have been completed and are awaiting analysis.

\section{SUMMARY}
\label{s:summary}
We have presented observations and analysis of the atmospheric transmission spectra of four transiting exoplanets using data taken with the HRS instrument on the HET.  We search for \ion{Na}{1} and \ion{K}{1} absorption in four targets.  Our target list includes HD 189733b, in which \citet{Redfield2008} detected \ion{Na}{1} absorption.  In an independent analysis of the same dataset, we confirm the \citet{Redfield2008} results.  We also tentatively confirm the results of \citet{Charbonneau2002} and \citet{Snellen2008} for \ion{Na}{1} absorption in HD 209458b.  Furthermore, we find a hint of \ion{Na}{1} absorption in HD 149026b but cannot make a conclusive detection because of the large random and systematic errors.  However, we suggest this is an excellent target for follow-up observations.  We do not find evidence of \ion{Na}{1} absorption in HD 147506b or \ion{K}{1} absorption in any of our targets.  Future work includes the analysis of a fifth target for which observations are complete, and searching for exospheric absorption in our targets.

\acknowledgements

Authors AGJ and SR acknowledge support by the National Science Foundation through Astronomy and Astrophysics Research Grant AST-0903573 (P.I.:  SR).  The Hobby-Eberly Telescope is a joint project of the University of Texas at Austin, the Pennsylvania State University, Stanford University, Ludwig-Maximilians-Universit\"{a}t M\"{u}nchen, and Georg-August-Universit\"{a}t G\"{o}ttingen and is named in honor of its principal benefactors, William P.~Hobby and Robert E.~Eberly.  This work made use of IDL, the Interactive Data Language\footnote{http://www.ittvis.com/language/en-us/productsservices/idl.aspx}; IRAF, the Image Reduction and Analysis Facility\footnote{http://iraf.noao.edu/}; the SIMBAD Database\footnote{http://simbad.u-strasbg.fr/simbad/}; the Exoplanet Data Explorer\footnote{http://exoplanets.org/}; and the Exoplanet Transit Database\footnote{http://var2.astro.cz/ETD/index.php}.

\bibliographystyle{apj}
\bibliography{refs}

\begin{thebibliography}{71}
\expandafter\ifx\csname natexlab\endcsname\relax\def\natexlab#1{#1}\fi

\bibitem[{{Agol} {et~al.}(2010){Agol}, {Cowan}, {Knutson}, {Deming}, {Steffen},
  {Henry}, \& {Charbonneau}}]{Agol2010}
{Agol}, E., {Cowan}, N.~B., {Knutson}, H.~A., {Deming}, D., {Steffen}, J.~H.,
  {Henry}, G.~W., \& {Charbonneau}, D. 2010, \apj, 721, 1861

\bibitem[{{Arai} {et~al.}(2010){Arai}, {Redfield}, \& {Koesterke}}]{Arai2010}
{Arai}, E., {Redfield}, S., \& {Koesterke}, L. 2010, in Bulletin of the
  American Astronomical Society, Vol.~42, American Astronomical Society Meeting
  Abstracts \#215, \#423.16

\bibitem[{{Bakos} {et~al.}(2010){Bakos}, {Torres}, {P{\'a}l}, {Hartman},
  {Kov{\'a}cs}, {Noyes}, {Latham}, {Sasselov}, {Sip{\H o}cz}, {Esquerdo},
  {Fischer}, {Johnson}, {Marcy}, {Butler}, {Isaacson}, {Howard}, {Vogt},
  {Kov{\'a}cs}, {Fernandez}, {Mo{\'o}r}, {Stefanik}, {L{\'a}z{\'a}r}, {Papp},
  \& {S{\'a}ri}}]{Bakos2010}
{Bakos}, G.~{\'A}., {Torres}, G., {P{\'a}l}, A., {Hartman}, J., {Kov{\'a}cs},
  G., {Noyes}, R.~W., {Latham}, D.~W., {Sasselov}, D.~D., {Sip{\H o}cz}, B.,
  {Esquerdo}, G.~A., {Fischer}, D.~A., {Johnson}, J.~A., {Marcy}, G.~W.,
  {Butler}, R.~P., {Isaacson}, H., {Howard}, A., {Vogt}, S., {Kov{\'a}cs}, G.,
  {Fernandez}, J., {Mo{\'o}r}, A., {Stefanik}, R.~P., {L{\'a}z{\'a}r}, J.,
  {Papp}, I., \& {S{\'a}ri}, P. 2010, \apj, 710, 1724

\bibitem[{{Ballester} {et~al.}(2007){Ballester}, {Sing}, \&
  {Herbert}}]{Ballester2007}
{Ballester}, G.~E., {Sing}, D.~K., \& {Herbert}, F. 2007, \nat, 445, 511

\bibitem[{{Barman}(2007)}]{Barman2007}
{Barman}, T. 2007, \apjl, 661, L191

\bibitem[{{Barman} {et~al.}(2001){Barman}, {Hauschildt}, \&
  {Allard}}]{Barman2001}
{Barman}, T.~S., {Hauschildt}, P.~H., \& {Allard}, F. 2001, \apj, 556, 885

\bibitem[{{Beaulieu} {et~al.}(2008){Beaulieu}, {Carey}, {Ribas}, \&
  {Tinetti}}]{Beaulieu2008}
{Beaulieu}, J.~P., {Carey}, S., {Ribas}, I., \& {Tinetti}, G. 2008, \apj, 677,
  1343

\bibitem[{{Beaulieu} {et~al.}(2010){Beaulieu}, {Kipping}, {Batista}, {Tinetti},
  {Ribas}, {Carey}, {Noriega-Crespo}, {Griffith}, {Campanella}, {Dong},
  {Tennyson}, {Barber}, {Deroo}, {Fossey}, {Liang}, {Swain}, {Yung}, \&
  {Allard}}]{Beaulieu2010}
{Beaulieu}, J.~P., {Kipping}, D.~M., {Batista}, V., {Tinetti}, G., {Ribas}, I.,
  {Carey}, S., {Noriega-Crespo}, J.~A., {Griffith}, C.~A., {Campanella}, G.,
  {Dong}, S., {Tennyson}, J., {Barber}, R.~J., {Deroo}, P., {Fossey}, S.~J.,
  {Liang}, D., {Swain}, M.~R., {Yung}, Y., \& {Allard}, N. 2010, \mnras, 409,
  963

\bibitem[{{Bouchy} {et~al.}(2005){Bouchy}, {Udry}, {Mayor}, {Moutou}, {Pont},
  {Iribarne}, {da Silva}, {Ilovaisky}, {Queloz}, {Santos}, {S{\'e}gransan}, \&
  {Zucker}}]{Bouchy2005}
{Bouchy}, F., {Udry}, S., {Mayor}, M., {Moutou}, C., {Pont}, F., {Iribarne},
  N., {da Silva}, R., {Ilovaisky}, S., {Queloz}, D., {Santos}, N.~C.,
  {S{\'e}gransan}, D., \& {Zucker}, S. 2005, \aap, 444, L15

\bibitem[{{Brown}(2001)}]{Brown2001}
{Brown}, T.~M. 2001, \apj, 553, 1006

\bibitem[{{Burrows} {et~al.}(2007){Burrows}, {Hubeny}, {Budaj}, {Knutson}, \&
  {Charbonneau}}]{Burrows2007}
{Burrows}, A., {Hubeny}, I., {Budaj}, J., {Knutson}, H.~A., \& {Charbonneau},
  D. 2007, \apjl, 668, L171

\bibitem[{{Butler} {et~al.}(2006){Butler}, {Wright}, {Marcy}, {Fischer},
  {Vogt}, {Tinney}, {Jones}, {Carter}, {Johnson}, {McCarthy}, \&
  {Penny}}]{Butler2006}
{Butler}, R.~P., {Wright}, J.~T., {Marcy}, G.~W., {Fischer}, D.~A., {Vogt},
  S.~S., {Tinney}, C.~G., {Jones}, H.~R.~A., {Carter}, B.~D., {Johnson}, J.~A.,
  {McCarthy}, C., \& {Penny}, A.~J. 2006, \apj, 646, 505

\bibitem[{{Carter} {et~al.}(2009){Carter}, {Winn}, {Gilliland}, \&
  {Holman}}]{Carter2009}
{Carter}, J.~A., {Winn}, J.~N., {Gilliland}, R., \& {Holman}, M.~J. 2009, \apj,
  696, 241

\bibitem[{{Castelli} \& {Kurucz}(2006)}]{CastelliKurucz2006}
{Castelli}, F., \& {Kurucz}, R.~L. 2006, \aap, 454, 333

\bibitem[{{Charbonneau} {et~al.}(2000){Charbonneau}, {Brown}, {Latham}, \&
  {Mayor}}]{Charbonneau2000}
{Charbonneau}, D., {Brown}, T.~M., {Latham}, D.~W., \& {Mayor}, M. 2000, \apjl,
  529, L45

\bibitem[{{Charbonneau} {et~al.}(2002){Charbonneau}, {Brown}, {Noyes}, \&
  {Gilliland}}]{Charbonneau2002}
{Charbonneau}, D., {Brown}, T.~M., {Noyes}, R.~W., \& {Gilliland}, R.~L. 2002,
  \apj, 568, 377

\bibitem[{{Charbonneau} {et~al.}(2008){Charbonneau}, {Knutson}, {Barman},
  {Allen}, {Mayor}, {Megeath}, {Queloz}, \& {Udry}}]{Charbonneau2008}
{Charbonneau}, D., {Knutson}, H.~A., {Barman}, T., {Allen}, L.~E., {Mayor}, M.,
  {Megeath}, S.~T., {Queloz}, D., \& {Udry}, S. 2008, \apj, 686, 1341

\bibitem[{{Colon} {et~al.}(2010){Colon}, {Ford}, {Redfield}, {Fortney},
  {Shabram}, {Deeg}, \& {Mahadevan}}]{Colon2010}
{Colon}, K.~D., {Ford}, E.~B., {Redfield}, S., {Fortney}, J.~J., {Shabram}, M.,
  {Deeg}, H.~J., \& {Mahadevan}, S. 2010, ArXiv e-prints, \#1008.4800

\bibitem[{{D{\'e}sert} {et~al.}(2008){D{\'e}sert}, {Vidal-Madjar}, {Lecavelier
  Des Etangs}, {Sing}, {Ehrenreich}, {H{\'e}brard}, \& {Ferlet}}]{Desert2008}
{D{\'e}sert}, J.-M., {Vidal-Madjar}, A., {Lecavelier Des Etangs}, A., {Sing},
  D., {Ehrenreich}, D., {H{\'e}brard}, G., \& {Ferlet}, R. 2008, \aap, 492, 585

\bibitem[{{Ehrenreich} {et~al.}(2007){Ehrenreich}, {H{\'e}brard}, {Lecavelier
  des Etangs}, {Sing}, {D{\'e}sert}, {Bouchy}, {Ferlet}, \&
  {Vidal-Madjar}}]{Ehrenreich2007}
{Ehrenreich}, D., {H{\'e}brard}, G., {Lecavelier des Etangs}, A., {Sing},
  D.~K., {D{\'e}sert}, J., {Bouchy}, F., {Ferlet}, R., \& {Vidal-Madjar}, A.
  2007, \apjl, 668, L179

\bibitem[{{Fitzpatrick} \& {Massa}(2005)}]{FM2005}
{Fitzpatrick}, E.~L., \& {Massa}, D. 2005, \aj, 129, 1642

\bibitem[{{Fortney} {et~al.}(2003){Fortney}, {Sudarsky}, {Hubeny}, {Cooper},
  {Hubbard}, {Burrows}, \& {Lunine}}]{Fortney2003}
{Fortney}, J.~J., {Sudarsky}, D., {Hubeny}, I., {Cooper}, C.~S., {Hubbard},
  W.~B., {Burrows}, A., \& {Lunine}, J.~I. 2003, \apj, 589, 615

\bibitem[{{Fossati} {et~al.}(2010){Fossati}, {Haswell}, {Froning}, {Hebb},
  {Holmes}, {Kolb}, {Helling}, {Carter}, {Wheatley}, {Collier Cameron},
  {Loeillet}, {Pollacco}, {Street}, {Stempels}, {Simpson}, {Udry}, {Joshi},
  {West}, {Skillen}, \& {Wilson}}]{Fossati2010}
{Fossati}, L., {Haswell}, C.~A., {Froning}, C.~S., {Hebb}, L., {Holmes}, S.,
  {Kolb}, U., {Helling}, C., {Carter}, A., {Wheatley}, P., {Collier Cameron},
  A., {Loeillet}, B., {Pollacco}, D., {Street}, R., {Stempels}, H.~C.,
  {Simpson}, E., {Udry}, S., {Joshi}, Y.~C., {West}, R.~G., {Skillen}, I., \&
  {Wilson}, D. 2010, \apjl, 714, L222

\bibitem[{{Gibson} {et~al.}(2011){Gibson}, {Pont}, \& {Aigrain}}]{Gibson2011}
{Gibson}, N.~P., {Pont}, F., \& {Aigrain}, S. 2011, \mnras, 411, 2199

\bibitem[{{Grillmair} {et~al.}(2008){Grillmair}, {Burrows}, {Charbonneau},
  {Armus}, {Stauffer}, {Meadows}, {van Cleve}, {von Braun}, \&
  {Levine}}]{Grillmair2008}
{Grillmair}, C.~J., {Burrows}, A., {Charbonneau}, D., {Armus}, L., {Stauffer},
  J., {Meadows}, V., {van Cleve}, J., {von Braun}, K., \& {Levine}, D. 2008,
  \nat, 456, 767

\bibitem[{{Grillmair} {et~al.}(2007){Grillmair}, {Charbonneau}, {Burrows},
  {Armus}, {Stauffer}, {Meadows}, {Van Cleve}, \& {Levine}}]{Grillmair2007}
{Grillmair}, C.~J., {Charbonneau}, D., {Burrows}, A., {Armus}, L., {Stauffer},
  J., {Meadows}, V., {Van Cleve}, J., \& {Levine}, D. 2007, \apjl, 658, L115

\bibitem[{{Harrington} {et~al.}(2007){Harrington}, {Luszcz}, {Seager},
  {Deming}, \& {Richardson}}]{Harrington2007}
{Harrington}, J., {Luszcz}, S., {Seager}, S., {Deming}, D., \& {Richardson},
  L.~J. 2007, \nat, 447, 691

\bibitem[{{Henry} {et~al.}(2000){Henry}, {Marcy}, {Butler}, \&
  {Vogt}}]{Henry2000}
{Henry}, G.~W., {Marcy}, G.~W., {Butler}, R.~P., \& {Vogt}, S.~S. 2000, \apjl,
  529, L41

\bibitem[{{Hubbard} {et~al.}(2001){Hubbard}, {Fortney}, {Lunine}, {Burrows},
  {Sudarsky}, \& {Pinto}}]{Hubbard2001}
{Hubbard}, W.~B., {Fortney}, J.~J., {Lunine}, J.~I., {Burrows}, A., {Sudarsky},
  D., \& {Pinto}, P. 2001, \apj, 560, 413

\bibitem[{{Hubeny} {et~al.}(2003){Hubeny}, {Burrows}, \&
  {Sudarsky}}]{Hubeny2003}
{Hubeny}, I., {Burrows}, A., \& {Sudarsky}, D. 2003, \apj, 594, 1011

\bibitem[{{Knutson} {et~al.}(2008){Knutson}, {Charbonneau}, {Allen}, {Burrows},
  \& {Megeath}}]{Knutson2008}
{Knutson}, H.~A., {Charbonneau}, D., {Allen}, L.~E., {Burrows}, A., \&
  {Megeath}, S.~T. 2008, \apj, 673, 526

\bibitem[{{Knutson} {et~al.}(2009){Knutson}, {Charbonneau}, {Cowan}, {Fortney},
  {Showman}, {Agol}, \& {Henry}}]{Knutson2009}
{Knutson}, H.~A., {Charbonneau}, D., {Cowan}, N.~B., {Fortney}, J.~J.,
  {Showman}, A.~P., {Agol}, E., \& {Henry}, G.~W. 2009, \apj, 703, 769

\bibitem[{{Knutson} {et~al.}(2010){Knutson}, {Howard}, \&
  {Isaacson}}]{Knutson2010}
{Knutson}, H.~A., {Howard}, A.~W., \& {Isaacson}, H. 2010, \apj, 720, 1569

\bibitem[{{Koesterke} {et~al.}(2008){Koesterke}, {Allende Prieto}, \&
  {Lambert}}]{Koesterke2008}
{Koesterke}, L., {Allende Prieto}, C., \& {Lambert}, D.~L. 2008, \apj, 680, 764

\bibitem[{{Latham} {et~al.}(1989){Latham}, {Stefanik}, {Mazeh}, {Mayor}, \&
  {Burki}}]{Latham1989}
{Latham}, D.~W., {Stefanik}, R.~P., {Mazeh}, T., {Mayor}, M., \& {Burki}, G.
  1989, \nat, 339, 38

\bibitem[{{Lecavelier Des Etangs} {et~al.}(2010){Lecavelier Des Etangs},
  {Ehrenreich}, {Vidal-Madjar}, {Ballester}, {D{\'e}sert}, {Ferlet},
  {H{\'e}brard}, {Sing}, {Tchakoumegni}, \& {Udry}}]{LecavelierDesEtangs2010}
{Lecavelier Des Etangs}, A., {Ehrenreich}, D., {Vidal-Madjar}, A., {Ballester},
  G.~E., {D{\'e}sert}, J., {Ferlet}, R., {H{\'e}brard}, G., {Sing}, D.~K.,
  {Tchakoumegni}, K., \& {Udry}, S. 2010, \aap, 514, A72

\bibitem[{{Lodders}(2003)}]{Lodders}
{Lodders}, K. 2003, \apj, 591, 1220

\bibitem[{{Markwardt}(2009)}]{Markwardt2009}
{Markwardt}, C.~B. 2009, in Astronomical Society of the Pacific Conference
  Series, Vol. 411, Astronomical Data Analysis Software and Systems XVIII, ed.
  {D.~A.~Bohlender, D.~Durand, \& P.~Dowler}, 251

\bibitem[{{Mayor} \& {Queloz}(1995)}]{MayorQueloz1995}
{Mayor}, M., \& {Queloz}, D. 1995, \nat, 378, 355

\bibitem[{{Mazeh} {et~al.}(2000){Mazeh}, {Naef}, {Torres}, {Latham}, {Mayor},
  {Beuzit}, {Brown}, {Buchhave}, {Burnet}, {Carney}, {Charbonneau}, {Drukier},
  {Laird}, {Pepe}, {Perrier}, {Queloz}, {Santos}, {Sivan}, {Udry}, \&
  {Zucker}}]{Mazeh2000}
{Mazeh}, T., {Naef}, D., {Torres}, G., {Latham}, D.~W., {Mayor}, M., {Beuzit},
  J., {Brown}, T.~M., {Buchhave}, L., {Burnet}, M., {Carney}, B.~W.,
  {Charbonneau}, D., {Drukier}, G.~A., {Laird}, J.~B., {Pepe}, F., {Perrier},
  C., {Queloz}, D., {Santos}, N.~C., {Sivan}, J., {Udry}, S., \& {Zucker}, S.
  2000, \apjl, 532, L55

\bibitem[{{P{\'a}l} {et~al.}(2010){P{\'a}l}, {Bakos}, {Torres}, {Noyes},
  {Fischer}, {Johnson}, {Henry}, {Butler}, {Marcy}, {Howard}, {Sip{\H o}cz},
  {Latham}, \& {Esquerdo}}]{Pal2010}
{P{\'a}l}, A., {Bakos}, G.~{\'A}., {Torres}, G., {Noyes}, R.~W., {Fischer},
  D.~A., {Johnson}, J.~A., {Henry}, G.~W., {Butler}, R.~P., {Marcy}, G.~W.,
  {Howard}, A.~W., {Sip{\H o}cz}, B., {Latham}, D.~W., \& {Esquerdo}, G.~A.
  2010, \mnras, 401, 2665

\bibitem[{{Pont} {et~al.}(2008){Pont}, {Knutson}, {Gilliland}, {Moutou}, \&
  {Charbonneau}}]{Pont2008}
{Pont}, F., {Knutson}, H., {Gilliland}, R.~L., {Moutou}, C., \& {Charbonneau},
  D. 2008, \mnras, 385, 109

\bibitem[{{Press} {et~al.}(1992){Press}, {Teukolsky}, {Vetterling}, \&
  {Flannery}}]{Press1992}
{Press}, W.~H., {Teukolsky}, S.~A., {Vetterling}, W.~T., \& {Flannery}, B.~P.
  1992, {Numerical recipes in C. The art of scientific computing} (Cambridge:
  University Press, 1992, 2nd ed.)

\bibitem[{{Rasio}(1994)}]{Rasio1994}
{Rasio}, F.~A. 1994, \apjl, 427, L107

\bibitem[{{Redfield} {et~al.}(2008){Redfield}, {Endl}, {Cochran}, \&
  {Koesterke}}]{Redfield2008}
{Redfield}, S., {Endl}, M., {Cochran}, W.~D., \& {Koesterke}, L. 2008, \apjl,
  673, L87

\bibitem[{{Sato} {et~al.}(2005){Sato}, {Fischer}, {Henry}, {Laughlin},
  {Butler}, {Marcy}, {Vogt}, {Bodenheimer}, {Ida}, {Toyota}, {Wolf}, {Valenti},
  {Boyd}, {Johnson}, {Wright}, {Ammons}, {Robinson}, {Strader}, {McCarthy},
  {Tah}, \& {Minniti}}]{Sato2005}
{Sato}, B., {Fischer}, D.~A., {Henry}, G.~W., {Laughlin}, G., {Butler}, R.~P.,
  {Marcy}, G.~W., {Vogt}, S.~S., {Bodenheimer}, P., {Ida}, S., {Toyota}, E.,
  {Wolf}, A., {Valenti}, J.~A., {Boyd}, L.~J., {Johnson}, J.~A., {Wright},
  J.~T., {Ammons}, M., {Robinson}, S., {Strader}, J., {McCarthy}, C., {Tah},
  K.~L., \& {Minniti}, D. 2005, \apj, 633, 465

\bibitem[{{Seager} \& {Deming}(2010)}]{SeagerDeming2010}
{Seager}, S., \& {Deming}, D. 2010, \araa, 48, 631

\bibitem[{{Seager} \& {Sasselov}(2000)}]{SeagerSasselov2000}
{Seager}, S., \& {Sasselov}, D.~D. 2000, \apj, 537, 916

\bibitem[{{Sing} {et~al.}(2011){Sing}, {D{\'e}sert}, {Fortney}, {Lecavelier Des
  Etangs}, {Ballester}, {Cepa}, {Ehrenreich}, {L{\'o}pez-Morales}, {Pont},
  {Shabram}, \& {Vidal-Madjar}}]{Sing2011}
{Sing}, D.~K., {D{\'e}sert}, J., {Fortney}, J.~J., {Lecavelier Des Etangs}, A.,
  {Ballester}, G.~E., {Cepa}, J., {Ehrenreich}, D., {L{\'o}pez-Morales}, M.,
  {Pont}, F., {Shabram}, M., \& {Vidal-Madjar}, A. 2011, \aap, 527, A73

\bibitem[{{Sing} {et~al.}(2008{\natexlab{a}}){Sing}, {Vidal-Madjar},
  {D{\'e}sert}, {Lecavelier des Etangs}, \& {Ballester}}]{Sing2008a}
{Sing}, D.~K., {Vidal-Madjar}, A., {D{\'e}sert}, J., {Lecavelier des Etangs},
  A., \& {Ballester}, G. 2008{\natexlab{a}}, \apj, 686, 658

\bibitem[{{Sing} {et~al.}(2008{\natexlab{b}}){Sing}, {Vidal-Madjar},
  {Lecavelier des Etangs}, {D{\'e}sert}, {Ballester}, \&
  {Ehrenreich}}]{Sing2008b}
{Sing}, D.~K., {Vidal-Madjar}, A., {Lecavelier des Etangs}, A., {D{\'e}sert},
  J.-M., {Ballester}, G., \& {Ehrenreich}, D. 2008{\natexlab{b}}, \apj, 686,
  667

\bibitem[{{Snellen} {et~al.}(2008){Snellen}, {Albrecht}, {de Mooij}, \& {Le
  Poole}}]{Snellen2008}
{Snellen}, I.~A.~G., {Albrecht}, S., {de Mooij}, E.~J.~W., \& {Le Poole}, R.~S.
  2008, \aap, 487, 357

\bibitem[{{Snellen} {et~al.}(2010){Snellen}, {de Kok}, {de Mooij}, \&
  {Albrecht}}]{Snellen2010}
{Snellen}, I.~A.~G., {de Kok}, R.~J., {de Mooij}, E.~J.~W., \& {Albrecht}, S.
  2010, \nat, 465, 1049

\bibitem[{{Spiegel} {et~al.}(2009){Spiegel}, {Silverio}, \&
  {Burrows}}]{Spiegel2009}
{Spiegel}, D.~S., {Silverio}, K., \& {Burrows}, A. 2009, \apj, 699, 1487

\bibitem[{{Swain} {et~al.}(2010){Swain}, {Deroo}, {Griffith}, {Tinetti},
  {Thatte}, {Vasisht}, {Chen}, {Bouwman}, {Crossfield}, {Angerhausen},
  {Afonso}, \& {Henning}}]{Swain2010}
{Swain}, M.~R., {Deroo}, P., {Griffith}, C.~A., {Tinetti}, G., {Thatte}, A.,
  {Vasisht}, G., {Chen}, P., {Bouwman}, J., {Crossfield}, I.~J., {Angerhausen},
  D., {Afonso}, C., \& {Henning}, T. 2010, \nat, 463, 637

\bibitem[{{Swain} {et~al.}(2009{\natexlab{a}}){Swain}, {Tinetti}, {Vasisht},
  {Deroo}, {Griffith}, {Bouwman}, {Chen}, {Yung}, {Burrows}, {Brown},
  {Matthews}, {Rowe}, {Kuschnig}, \& {Angerhausen}}]{Swain2009b}
{Swain}, M.~R., {Tinetti}, G., {Vasisht}, G., {Deroo}, P., {Griffith}, C.,
  {Bouwman}, J., {Chen}, P., {Yung}, Y., {Burrows}, A., {Brown}, L.~R.,
  {Matthews}, J., {Rowe}, J.~F., {Kuschnig}, R., \& {Angerhausen}, D.
  2009{\natexlab{a}}, \apj, 704, 1616

\bibitem[{{Swain} {et~al.}(2008){Swain}, {Vasisht}, \& {Tinetti}}]{Swain2008}
{Swain}, M.~R., {Vasisht}, G., \& {Tinetti}, G. 2008, \nat, 452, 329

\bibitem[{{Swain} {et~al.}(2009{\natexlab{b}}){Swain}, {Vasisht}, {Tinetti},
  {Bouwman}, {Chen}, {Yung}, {Deming}, \& {Deroo}}]{Swain2009a}
{Swain}, M.~R., {Vasisht}, G., {Tinetti}, G., {Bouwman}, J., {Chen}, P.,
  {Yung}, Y., {Deming}, D., \& {Deroo}, P. 2009{\natexlab{b}}, \apjl, 690, L114

\bibitem[{{Tinetti} {et~al.}(2007){Tinetti}, {Vidal-Madjar}, {Liang},
  {Beaulieu}, {Yung}, {Carey}, {Barber}, {Tennyson}, {Ribas}, {Allard},
  {Ballester}, {Sing}, \& {Selsis}}]{Tinetti2007}
{Tinetti}, G., {Vidal-Madjar}, A., {Liang}, M., {Beaulieu}, J., {Yung}, Y.,
  {Carey}, S., {Barber}, R.~J., {Tennyson}, J., {Ribas}, I., {Allard}, N.,
  {Ballester}, G.~E., {Sing}, D.~K., \& {Selsis}, F. 2007, \nat, 448, 169

\bibitem[{{Torres} {et~al.}(2008){Torres}, {Winn}, \& {Holman}}]{Torres2008}
{Torres}, G., {Winn}, J.~N., \& {Holman}, M.~J. 2008, \apj, 677, 1324

\bibitem[{{Triaud} {et~al.}(2009){Triaud}, {Queloz}, {Bouchy}, {Moutou},
  {Collier Cameron}, {Claret}, {Barge}, {Benz}, {Deleuil}, {Guillot},
  {H{\'e}brard}, {Lecavelier Des {\'E}tangs}, {Lovis}, {Mayor}, {Pepe}, \&
  {Udry}}]{Triaud2009}
{Triaud}, A.~H.~M.~J., {Queloz}, D., {Bouchy}, F., {Moutou}, C., {Collier
  Cameron}, A., {Claret}, A., {Barge}, P., {Benz}, W., {Deleuil}, M.,
  {Guillot}, T., {H{\'e}brard}, G., {Lecavelier Des {\'E}tangs}, A., {Lovis},
  C., {Mayor}, M., {Pepe}, F., \& {Udry}, S. 2009, \aap, 506, 377

\bibitem[{{Vidal-Madjar} {et~al.}(2004){Vidal-Madjar}, {D{\'e}sert},
  {Lecavelier des Etangs}, {H{\'e}brard}, {Ballester}, {Ehrenreich}, {Ferlet},
  {McConnell}, {Mayor}, \& {Parkinson}}]{VidalMadjar2004}
{Vidal-Madjar}, A., {D{\'e}sert}, J., {Lecavelier des Etangs}, A.,
  {H{\'e}brard}, G., {Ballester}, G.~E., {Ehrenreich}, D., {Ferlet}, R.,
  {McConnell}, J.~C., {Mayor}, M., \& {Parkinson}, C.~D. 2004, \apjl, 604, L69

\bibitem[{{Vidal-Madjar} {et~al.}(2003){Vidal-Madjar}, {Lecavelier des Etangs},
  {D{\'e}sert}, {Ballester}, {Ferlet}, {H{\'e}brard}, \&
  {Mayor}}]{VidalMadjar2003}
{Vidal-Madjar}, A., {Lecavelier des Etangs}, A., {D{\'e}sert}, J., {Ballester},
  G.~E., {Ferlet}, R., {H{\'e}brard}, G., \& {Mayor}, M. 2003, \nat, 422, 143

\bibitem[{{Vidal-Madjar} {et~al.}(2011){Vidal-Madjar}, {Sing}, {Lecavelier Des
  Etangs}, {Ferlet}, {D{\'e}sert}, {H{\'e}brard}, {Boisse}, {Ehrenreich}, \&
  {Moutou}}]{VidalMadjar2011}
{Vidal-Madjar}, A., {Sing}, D.~K., {Lecavelier Des Etangs}, A., {Ferlet}, R.,
  {D{\'e}sert}, J.-M., {H{\'e}brard}, G., {Boisse}, I., {Ehrenreich}, D., \&
  {Moutou}, C. 2011, \aap, 527, A110+

\bibitem[{{Wall} \& {Jenkins}(2003)}]{Wall2003}
{Wall}, J.~V., \& {Jenkins}, C.~R. 2003, {Practical Statistics for Astronomers}
  (Princeton Series in Astrophysics)

\bibitem[{{Winn} {et~al.}(2007){Winn}, {Johnson}, {Peek}, {Marcy}, {Bakos},
  {Enya}, {Narita}, {Suto}, {Turner}, \& {Vogt}}]{Winn2007}
{Winn}, J.~N., {Johnson}, J.~A., {Peek}, K.~M.~G., {Marcy}, G.~W., {Bakos},
  G.~{\'A}., {Enya}, K., {Narita}, N., {Suto}, Y., {Turner}, E.~L., \& {Vogt},
  S.~S. 2007, \apjl, 665, L167

\bibitem[{{Winn} {et~al.}(2005){Winn}, {Noyes}, {Holman}, {Charbonneau},
  {Ohta}, {Taruya}, {Suto}, {Narita}, {Turner}, {Johnson}, {Marcy}, {Butler},
  \& {Vogt}}]{Winn2005}
{Winn}, J.~N., {Noyes}, R.~W., {Holman}, M.~J., {Charbonneau}, D., {Ohta}, Y.,
  {Taruya}, A., {Suto}, Y., {Narita}, N., {Turner}, E.~L., {Johnson}, J.~A.,
  {Marcy}, G.~W., {Butler}, R.~P., \& {Vogt}, S.~S. 2005, \apj, 631, 1215

\bibitem[{{Winn} {et~al.}(2004){Winn}, {Suto}, {Turner}, {Narita}, {Frye},
  {Aoki}, {Sato}, \& {Yamada}}]{Winn2004}
{Winn}, J.~N., {Suto}, Y., {Turner}, E.~L., {Narita}, N., {Frye}, B.~L.,
  {Aoki}, W., {Sato}, B., \& {Yamada}, T. 2004, \pasj, 56, 655

\bibitem[{{Wolf} {et~al.}(2007){Wolf}, {Laughlin}, {Henry}, {Fischer}, {Marcy},
  {Butler}, \& {Vogt}}]{Wolf2007}
{Wolf}, A.~S., {Laughlin}, G., {Henry}, G.~W., {Fischer}, D.~A., {Marcy}, G.,
  {Butler}, P., \& {Vogt}, S. 2007, \apj, 667, 549

\bibitem[{{Wolszczan} \& {Frail}(1992)}]{WolszczanFrail1992}
{Wolszczan}, A., \& {Frail}, D.~A. 1992, \nat, 355, 145

\bibitem[{{Wood} {et~al.}(2011){Wood}, {Maxted}, {Smalley}, \&
  {Iro}}]{Wood2011}
{Wood}, P.~L., {Maxted}, P.~F.~L., {Smalley}, B., \& {Iro}, N. 2011, \mnras,
  171

\end{thebibliography}

\clearpage \clearpage

\clearpage \clearpage

\clearpage \clearpage

\clearpage \clearpage

\clearpage \clearpage

\clearpage \clearpage

\clearpage \clearpage

\clearpage \clearpage

\clearpage \clearpage

\clearpage \clearpage

\begin{deluxetable}{cccccccccc}
\tablecolumns{10}
\tablewidth{0pc}
\rotate
\tabletypesize{\scriptsize}
\tablecaption{Observations\label{table:observations}}
\tablehead{\colhead{Target} & \multicolumn{2}{c}{In Transit\tablenotemark{a}} & \multicolumn{2}{c}{Out of Transit\tablenotemark{a}} & \multicolumn{2}{c}{Ingress/Egress\tablenotemark{a}} & \colhead{Telluric} & \multicolumn{2}{c}{Observations} \\
\colhead{} & \colhead {\# of Obs.} & \colhead{Time (s)\tablenotemark{b}} & \colhead {\# of Obs.} & \colhead{Time (s)\tablenotemark{b}} & \colhead{\# of Obs.} & \colhead{Time (s)\tablenotemark{b}} & \colhead{} & \colhead{\# of Obs.} & \colhead{$<$Time$>$ (s)\tablenotemark{b}}}
\startdata
HD 147506b\tablenotemark{c} & 28 & 16508 & 84 & 49687 & 5 & 3000 & HD 156729\tablenotemark{d} & 34 & 567 \\
HD 149026b & 22 (1) & 12718 (600) & 90 & 52734 & 5 & 3000 & HD 156729\tablenotemark{d} & 34 & 567 \\
HD 189733b & 38 (3) & 20140 (1640) & 173 (1) & 79011 ($\sim$0) & 15 (1) & 8243 (600) & HD 196867 & 45 & 291 \\
HD 209458b & 29 (6) & 17310 (3510) & 72 (3) & 43175 (1800) & 4 & 2400 & HD 218045 & 21 & 80 \\
\enddata
\tablenotetext{a}{Numbers shown are all observations taken, while the numbers in parentheses indicate observations that are discarded for various reasons, primary cloud-induced low S/N, but also various reduction artifacts.  The ``In Transit" and ``"Out of Transit" columns are mutually exclusive, defined by observation midpoint, while the ``Ingress/Egress" observations are a subset of the ``In Transit" observations, again defined by observation midpoint.  For our \ion{Na}{1} analysis, one additional 600 s out-of-transit observation is discarded for HD 189733b.  For our \ion{Ca}{1} analysis, one additional 443 s in-transit observation is discarded for HD 149026b.  Finally, for HD 147506b, several observations are discarded for \ion{K}{1}:  3 in-transit observations totaling 1800 s, and 6 out-of-transit observations totaling 3600 s.}
\tablenotetext{b}{Times for primary science target exposures are total times, while telluric exposure times are average times, as the telluric observations are always used independently and are never coadded.}
\tablenotetext{c}{HD 147506b is also known as HAT-P-2.}
\tablenotetext{d}{HD 147506b and HD 149026b use the same telluric.  The values in the table this telluric represent the same observations, i.e., the 34 observations listed do not distinguish between which primary science target was originally associated with the telluric observation.}
\end{deluxetable}

\begin{deluxetable}{cccccccc}
\tablecolumns{8}
\tablewidth{0pc}
\rotate
\tabletypesize{\scriptsize}
\tablecaption{System and Orbital Parameters of Targets\label{table:systemparams}}
\tablehead{\colhead{Target} & \colhead{Reference Epoch} & \colhead{Period} & \colhead{Transit Duration} & \colhead{$a$/R$_{\rm *}$} & \colhead{R$_{\rm P}$/R$_{\rm *}$} & \colhead{Inclination} & \colhead{References}\\
\colhead{} & \colhead{(HJD$-2400000$)} & \colhead{(days)} & \colhead{(minutes)} & \colhead{} & \colhead{} & \colhead{($^{\circ}$)}}
\startdata
HD 147506b & $54387.4937\pm0.00074$      & $5.633473\pm0.0000061$     & $257.76\pm1.87$ & $8.99^{+0.39}_{-0.41}$ & $0.07227\pm0.00061$ & $86.7^{+1.12}_{-0.87}$ & 1, 2, 3, 4\\
HD 149026b & $54456.7876\pm0.00014$      & $2.87589\pm0.00001$            & $194.4\pm8.7$      & $7.11^{+0.03}_{-0.81}$  & $0.0491^{+0.0018}_{-0.0005}$ & $90.0^{+0.0}_{-3.0}$ & 3, 5, 6\\
HD 189733b & $54279.43671\pm0.000015$ & $2.2185757\pm0.00000015$ & $106.1\pm0.7$      & $8.81\pm0.06$              & $0.15436\pm0.00022$ & $85.58\pm0.06$ & 3, 7, 8, 9 \\
HD 209458b & $52854.825\pm0.000135$      & $3.5247455\pm0.00000018$ & $184.2\pm1.6$      & $8.76\pm0.04$              & $0.12086\pm0.0001$ & $86.71\pm0.05$ & 3, 10 \\
\enddata
\tablerefs{(1) \citet{Pal2010}.  (2) \citet{Bakos2010}. (3) \citet{Torres2008}.  (4) \citet{Winn2007}.  (5) \citet{Carter2009}.  (6) \citet{Wolf2007}.  (7) \citet{Agol2010}.  (8) \citet{Triaud2009}.  (9) \citet{Bouchy2005}.  (10) \citet{Winn2005}.  This table also made use of the Exoplanet Data Explorer (http://exoplanets.org) and the Exoplanet Transit Database (http://var2.astro.cz/ETD/index.php).}
\end{deluxetable}

\begin{deluxetable}{cccc}
\tablecolumns{4}
\tablewidth{0pc}
\tabletypesize{\small}
\tablecaption{Absorption Line Data\label{table:absorptionlines}}
\tablehead{\colhead{Element} & \colhead{Wavelength(s) (\AA{})} & \colhead{Bandwidth (\AA{})\tablenotemark{a}} & \colhead{Use}}
\startdata
\ion{Na}{1} & 5889.9510, 5895.9242 & 12 & Alkali metal resonance line \\
\ion{K}{1} & 7698.9645 & 8 & Alkali metal resonance line \\
\ion{Ca}{1} & 6122.2217 & 8 & Control line \\
\enddata
\tablenotetext{a}{The bandwidth over which integration is performed.  The band is centered at the wavelength in the first column, or in the case of multiple lines for \ion{Na}{1}, their average.  We also use reference bands of equal bandwidth immediately to the blue and red of these bands.}
\end{deluxetable}

\begin{deluxetable}{crrr}
\tablecolumns{4}
\tablewidth{0pc}
\tabletypesize{\small}
\tablecaption{Absorption Results\label{table:results}}
\tablehead{\colhead{Star} & \colhead{\ion{Na}{1}}  & \colhead{\ion{Ca}{1}}  & \colhead{\ion{K}{1}}}
\startdata
HD 147506b & $(1.88\pm1.18)\times10^{-4}$ & $(-0.078\pm0.74)\times10^{-4}$ & $(2.17\pm1.02)\times10^{-4}$ \\
HD 149026b & $(-3.16\pm2.06)\times10^{-4}$ & $(1.15\pm0.78)\times10^{-4}$ &  $(-0.71\pm1.09)\times10^{-4}$ \\
HD 189733b & $(-5.26\pm1.69)\times10^{-4}$ & $(0.72\pm0.63)\times10^{-4}$ & $(0.77\pm1.04)\times10^{-4}$ \\
HD 209458b & $(-2.63\pm0.62)\times10^{-4}$ & $(-0.36\pm0.36)\times10^{-4}$ & $(-0.89\pm0.67)\times10^{-4}$ \\
\enddata
\end{deluxetable}

\clearpage \clearpage

\end{document}